\documentclass[12pt,letterpaper]{article}
\pdfoutput=1

\usepackage{amsmath,amssymb,calc, amsthm,bbm, epsfig,psfrag, mathtools,comment}
\usepackage{mathrsfs}
\usepackage{graphicx, enumerate}
\usepackage{float}

\usepackage[numbers,sort&compress]{natbib}

\usepackage[dvipsnames]{xcolor}
\usepackage{tikz}
\usepackage{tikz-cd}
\usepackage{tensor}
\newcommand{\ul}{\underline}

\usepackage[utf8]{inputenc} 
\definecolor{azure}{rgb}{0.0, 0.5, 1.0}
\definecolor{darkblue}{rgb}{0.15,0.35,0.7}
\definecolor{reddish}{rgb}{0.65, 0.2, 0.2}
\definecolor{brandeisblue}{rgb}{0.0, 0.44, 1.0}
\definecolor{ceruleanblue}{rgb}{0.16, 0.32, 0.75}
\definecolor{indigo(dye)}{rgb}{0.0, 0.25, 0.42}
\usepackage[linktocpage=true]{hyperref}
\hypersetup{
colorlinks=true,
citecolor=ceruleanblue,
linkcolor=ceruleanblue,
urlcolor=ceruleanblue,
pdfauthor={},
pdftitle={},
pdfsubject={}
}

\newcommand{\overbar}[1]{\mkern 1.5mu\overline{\mkern-1.5mu#1\mkern-1.5mu}\mkern 1.5mu}

\newcommand{\TT}{T\overbar{T}}

\DeclareMathOperator{\tr}{\text{tr}}
\DeclareMathOperator{\str}{\text{str}}

\DeclareFontEncoding{LS1}{}{}
\DeclareFontSubstitution{LS1}{stix}{m}{n}
\DeclareSymbolFont{stixsymbols}{LS1}{stixscr}{m}{n}
\SetSymbolFont{stixsymbols}{bold}{LS1}{stixscr}{b}{n}
\DeclareMathSymbol{\kay}{\mathalpha}{stixsymbols}{"6B}
\DeclareMathSymbol{\hay}{\mathalpha}{stixsymbols}{"68}

\makeatletter
\renewcommand\section{\@startsection {section}{1}{\z@}%
                               {-3.5ex \@plus -1ex \@minus -.2ex}
                               {2.3ex \@plus.2ex}%
                               {\normalfont\large\bfseries}}
\renewcommand\subsection{\@startsection{subsection}{2}{\z@}%
                                 {-3.25ex\@plus -1ex \@minus -.2ex}%
                                 {1.5ex \@plus .2ex}%
                                 {\normalfont\bfseries}}
\makeatother

\makeatletter
\newcommand*\bigcdot{\mathpalette\bigcdot@{.5}}
\newcommand*\bigcdot@[2]{\mathbin{\vcenter{\hbox{\scalebox{#2}{$\m@th#1\bullet$}}}}}
\makeatother
\newcommand{\deq}{\stackrel{\bigcdot}{=}}

\newfont{\goth}{ygoth.tfm scaled 1200}                   

\numberwithin{equation}{section}

\begin{document}
\begin{titlepage}
\begin{flushright}
\today
\end{flushright}
\vspace{5mm}

\begin{center}
{\Large \bf 
Auxiliary Field Deformations of (Semi-)Symmetric Space Sigma Models}
\end{center}

\begin{center}

{\bf
Daniele Bielli${}^{a}$,
Christian Ferko${}^{b,c}$,
Liam Smith${}^{d}$,\\
Gabriele Tartaglino-Mazzucchelli${}^{d}$
} \\
\vspace{5mm}

\footnotesize{
${}^{a}$
{\it 
High Energy Physics Research Unit, Faculty of Science \\ 
Chulalongkorn University, Bangkok 10330, Thailand
}
 \\~\\
${}^{b}$
{\it 
Department of Physics, Northeastern University, Boston, MA 02115, USA
}
 \\~\\
${}^{c}$
{\it 
The NSF Institute for Artificial Intelligence
and Fundamental Interactions
}
  \\~\\
${}^{d}$
{\it 
School of Mathematics and Physics, University of Queensland,
\\
 St Lucia, Brisbane, Queensland 4072, Australia}
}
\vspace{2mm}
~\\
\texttt{d.bielli4@gmail.com,
c.ferko@northeastern.edu,
liam.smith1@uq.net.au,
g.tartaglino-mazzucchelli@uq.edu.au
}\\
\vspace{2mm}

\end{center}

\begin{abstract}
\baselineskip=14pt

We generalize the auxiliary field deformations of the principal chiral model (PCM) introduced in \cite{Ferko:2024ali} and \cite{Bielli:2024ach} to sigma models whose target manifolds are symmetric or semi-symmetric spaces, including a Wess-Zumino term in the latter case. This gives rise to a new infinite family of classically integrable $\mathbb{Z}_2$ and $\mathbb{Z}_4$ coset models of the form which are of interest in applications of integrability to worldsheet string theory and holography. We demonstrate that every theory in this infinite class admits a zero-curvature representation for its equations of motion by exhibiting a Lax connection.  

\end{abstract}
\vspace{5mm}

\vfill
\end{titlepage}


\renewcommand{\thefootnote}{\arabic{footnote}}
\setcounter{footnote}{0}

\tableofcontents{}
\vspace{1cm}
\bigskip\hrule

\section{Introduction}\label{sec:intro}

Integrable sigma models in two space-time dimensions play a significant role in several areas of theoretical and mathematical physics. Their applications range from statistical mechanics to condensed matter physics, to areas of mathematics such as representation theory and algebra, and, importantly, to models of quantum gravity. In fact, sigma models are the defining playground of string theory. Their integrable versions have dominated part of the research on AdS/CFT dualities, which has played a key role in enhancing our (non-perturbative) understanding of both quantum fields and gravity theories. Many of the string backgrounds relevant for the AdS/CFT duality, such as ${\rm AdS}_5 \times {\rm S}^5$, ${\rm AdS}_4 \times {\rm \mathbb {CP}}^4$, ${\rm AdS}_3\times {\rm S}^3 \times {\rm T}^4$, ${\rm AdS}_3\times {\rm S}^3 \times {\rm S}^3\times {\rm S}^1$, and other examples and their deformations, are described by integrable sigma models. This fact has helped advance the analysis of several observables and the matching with dual gauge theories. Integrability on both sides of the dualities has led to solving the spectrum and the associated dilatation operator in $4d$ ${\mathcal{N}}=4$ super-Yang-Mills or ABJM $3d$ supersymmetric Chern-Simons theory, and also led to a wealth of information on dual $2d$ CFTs which remain an elusive subject in the ${\rm AdS}_3/{\rm CFT}_2$ case. We refer the reader to \cite{Beisert:2010jr,Demulder:2023bux} for reviews on integrability in AdS/CFT and related subjects.

One of the means to advance research on integrable sigma models has focused on the development of a novel understanding of universal classes of integrable deformations that can be applied to these theories. These include various examples such as Yang-Baxter deformations \cite{Klimcik:2002zj,Klimcik:2008eq}, 
$\lambda$ deformations \cite{Sfetsos:2013wia}, and classes of deformations that emerge from target space dualities, such as (fermionic) T-duality, Poisson-Lie duality, and more generally, deformations related to double field theory and the action of $O(D,D)$ transformations --- see the reviews \cite{Hoare:2021dix,Borsato:2023dis} for references on these active research topics.

Recently, a new family of integrable PCM-like theories, known as auxiliary field sigma models (AFSM), has been introduced \cite{Ferko:2024ali}. The construction was motivated by extending, in a systematic way, other examples of universal integrable deformations driven by operators constructed from the energy-momentum tensor. The most famous deformation of this kind is driven by the $\TT$ operator of \cite{Zamolodchikov:2004ce,Cavaglia:2016oda,Smirnov:2016lqw}, which is well-known to preserve integrability. Another example that has attracted attention in the last two years is the root-$\TT$ deformation \cite{Ferko:2022cix,Conti:2022egv,Babaei-Aghbolagh:2022uij,Babaei-Aghbolagh:2022leo}, which was also shown to preserve integrability \cite{Borsato:2022tmu} for a large class of integrable sigma models such as the principal chiral model (PCM) with and without Wess-Zumino term, symmetric space sigma models (SSSM), and semi-symmetric space sigma models (sSSSM) also deformed with the Wess-Zumino (WZ) term first introduced in \cite{Cagnazzo:2012se}. 

Inspired by a remarkable relationship between duality-invariant electrodynamics in four spacetime dimensions formulated in terms of Ivanov-Zupnik auxiliary fields \cite{Ivanov:2002ab,Ivanov:2003uj} and $\TT$-like deformations driven by arbitrary functions of the energy-momentum tensor $f(T_{\mu\nu})$ as established in \cite{Ferko:2023wyi}, in \cite{Ferko:2024ali} it was shown how arbitrary $\TT$-like deformations of the $2d$ PCM could be described by a single, generic function $E(\nu_2)$ of one variable constructed out of a combination ($\nu_2=\tr[v_+v_+]\tr[v_-v_-]$) of a vector auxiliary field $v_\alpha$. Shortly after, in \cite{Bielli:2024ach}, we extended this construction by promoting the interaction function $E ( \nu_2 )$ to a function $E ( \nu_2, \ldots, \nu_N )$ of several variables. Remarkably, all these deformed theories preserve the integrability of the PCM. Moreover, in \cite{Bielli:2024ach} we argued that the family of integrable deformations parametrised by $E ( \nu_2, \ldots, \nu_N )$ includes deformations of the PCM by both the stress tensor and higher-spin conserved currents, such as the Smirnov-Zamolodchikov higher-spin deformations of \cite{Smirnov:2016lqw}. In a systematic effort to merge these auxiliary field deformations with other types of deformations mentioned above, recently, we also investigated the role of (non-Abelian) T-duality and $\TT$ \cite{Bielli:2024khq}, and we extended the AFSM to include Yang-Baxter deformations \cite{Bielli:2024fnp}. Further results on AFSM were recently obtained by Fukushima and Yoshida in \cite{Fukushima:2024nxm}, where the $2d$ AFSM was realized in terms of the $4d$ Chern-Simons theory of \cite{Costello:2019tri,nekrasov_thesis,Costello:2013zra} coupled to auxiliary fields. 

We refer the reader to our recent works \cite{Ferko:2024ali,Bielli:2024khq,Bielli:2024fnp,Bielli:2024ach} for more motivations and references related to the AFSM and further context for this line of research. Towards extending the classification of this new large class of integrable deformations, the scope of this paper is simple: show that auxiliary field deformations also exist for SSSM and sSSSM. We will see that the construction closely follows all previous examples and that the coupling to auxiliary fields preserves integrability of the original sigma models.

Our paper is organised as follows. In Section \ref{sec:review}, we give a rapid overview of (higher-spin) auxiliary field deformations of the PCM, with and without Wess-Zumino term. Section \ref{sec:SSSM} then generalizes this auxiliary field construction to sigma models whose target space is an arbitrary symmetric coset, and demonstrates that all of these deformed models are classically integrable by exhibiting a Lax representation for their equations of motion. In Section \ref{sec:sSSSM}, we further generalize this construction to sigma models on semi-symmetric spaces (allowing for the possible inclusion of a WZ term) and again provide a zero-curvature representation for the equations of motion in this setting. Section \ref{sec:conclusion} summarizes our results and presents some interesting questions for future investigation. We have relegated the details of certain technical computations to Appendices \ref{app:SSSM} and \ref{app:sSSSM}.

\enlargethispage{\baselineskip}

\textbf{Note added}. On the same day on which this work first appeared on the arXiv, the interesting related article \cite{Cesaro:2024ipq} also appeared with overlapping but complementary results. In particular, the work \cite{Cesaro:2024ipq} generalizes the original spin-$2$ auxiliary field deformations of the principal chiral model, first presented in \cite{Ferko:2024ali}, to the setting of $\mathbb{Z}_N$ coset models, which includes symmetric and semi-symmetric space models as certain special cases. In contrast, the main result of the present article is to generalize the higher-spin auxiliary field deformations of \cite{Bielli:2024ach} to the case of semi-symmetric space sigma models (or $\mathbb{Z}_4$ cosets), which can admit non-trivial WZ terms. Therefore, the results of this paper are both more general (as we allow spin-$n$ auxiliary field deformations for any $n \geq 2$) and less general (as we focus only on $\mathbb{Z}_2$ and $\mathbb{Z}_4$ cosets, but not $\mathbb{Z}_N$ cosets) than those of \cite{Cesaro:2024ipq}.

\section{Review of Auxiliary Field Deformations}\label{sec:review}

In this Section, we will overview the essential aspects of auxiliary field deformations of $2d$ sigma models, primarily following \cite{Ferko:2024ali,Bielli:2024ach}. We focus only on those elements which will be important for the later generalization to symmetric and semi-symmetric spaces and will not, for instance, mention connections to T-duality \cite{Bielli:2024khq}, Yang-Baxter deformations \cite{Bielli:2024fnp}, or $4d$ Chern-Simons theory \cite{Fukushima:2024nxm}; we refer the reader to the cited works for these applications.

\subsection{Higher-Spin Auxiliary Field Sigma Model}\label{sec:afsm_review}

The simplest seed theory which one can use as input for auxiliary field deformations is the principal chiral model whose target space is a Lie group $G$ with semi-simple Lie algebra $\mathfrak{g}$. We will refer to the theory obtained by deforming the PCM through couplings to auxiliary fields in this way as the higher-spin auxiliary field sigma model, or more commonly, simply as the auxiliary field sigma model (AFSM). 

As is conventional, we will denote the fundamental group-valued field of the PCM as $g ( \sigma, \tau )$, which is a map from the $2d$ worldsheet $\Sigma$ into $G$. A distinguished Lie algebra valued $1$-form obtained from $g$ is the left-invariant Maurer-Cartan form
\begin{align}
    j = g^{-1} d g \, ,
\end{align}
whose pull-back to the worldsheet will be written as
\begin{align}\label{mc_defn}
    j_\alpha = g^{-1} \partial_\alpha g \, .
\end{align}
We will always use early lowercase Greek letters like $\alpha$, $\beta$ to refer to indices on $\Sigma$. As a consequence of its definition (\ref{mc_defn}), the Maurer-Cartan form satisfies the flatness condition
\begin{align}\label{mc_identity}
    \partial_\alpha j_\beta - \partial_\beta j_\alpha + [ j_\alpha, j_\beta ] = 0 \, .
\end{align}
Rather than using the usual worldsheet coordinates $\sigma^\alpha = ( \sigma, \tau )$, it will often be convenient to convert to light-cone coordinates
\begin{align}
    \sigma^{\pm} = \frac{1}{2} \left( \tau \pm \sigma \right) \, ,
\end{align}
in terms of which the Lagrangian for the PCM takes the simple form
\begin{align}\label{pcm_lc}
    \mathcal{L}_{\text{PCM}} = - \frac{1}{2} \tr \left( j_+ j_- \right) \, .
\end{align}
Here we write $\tr$ for a trace in the appropriate matrix representation for $j_{\pm} \in \mathfrak{g}$. In this article, we always assume that the worldsheet $\Sigma$ has the topology of either the plane or the cylinder, and that it is equipped with the flat Minkowski metric $\eta_{\alpha \beta}$ whose components are
\begin{align}\label{minkowski_defn}
    \eta_{+ -} = \eta_{- +} = - 2 \, ,
\end{align}
in light-cone coordinates. Using (\ref{minkowski_defn}), the PCM Lagrangian (\ref{pcm_lc}) may also be written as
\begin{align}
    \mathcal{L}_{\text{PCM}} = \frac{1}{2} \eta^{\alpha \beta} \tr \left( j_\alpha j_\beta \right) \, .
\end{align}
Our entire discussion of the PCM, and of its coupling to auxiliary fields, could be alternatively phrased in terms of the right-invariant Maurer-Cartan form $\tilde{j} = - ( d g ) g^{-1}$, and its pull-back $\tilde{j}_\alpha$, without changing any of the essential physical features. However, for definiteness, we will focus on the presentation in terms of the left-invariant form $j_\alpha$.

To pass from the undeformed PCM to the deformed AFSM, one introduces additional $\mathfrak{g}$-valued fields $v_{\pm}$ which couple to the Maurer-Cartan form $j_\alpha$ in a prescribed way which preserves classical integrability. In particular, we first define a collection of scalars
\begin{align}\label{scalars_defn}
    \nu_n = \tr ( v_+^n ) \tr ( v_-^n ) \, , \qquad n = 2 , \ldots , N \, ,
\end{align}
where we assume that the fields $v_{\pm} \in \mathfrak{g}$ are realized using an $N \times N$ matrix representation of the Lie algebra $\mathfrak{g}$. Let $T_A$ be a collection of generators for $\mathfrak{g}$, which we will always label with capital early Latin letters to distinguish these indices from worldsheet indices. Likewise, we will expand any Lie algebra valued quantity $X \in \mathfrak{g}$ in generators using the notation $X = X^A T_A$. Because $\mathfrak{g}$ is semi-simple, one has 
\begin{align}
    \tr ( v_{\pm} ) = \tr \left( v_{\pm}^A T_A \right) = 0 \, ,
\end{align}
since $\tr ( T_A ) = 0$ for each $A$. There are therefore at most $N - 1$ additional functionally independent traces $\tr ( v_{\alpha}^n )$ for $n = 2 , \ldots, N$ and for each choice of the index $\alpha \in \{ + , - \}$. Higher traces $\tr ( v_{\pm}^n )$ for $n > N$ can be expressed in terms of the lower traces using identities which follow from the Cayley-Hamilton theorem, which is why we need not consider any additional scalars $\nu_n$ for $n > N$, as these may be written in terms of the variables (\ref{scalars_defn}). 
The number of independent $\nu_n$ is further constrained when $N$ is bigger than the rank of the algebra $\mathfrak{g}$. In fact, $\nu_n$ is given by the product of two symmetric traces of order $n$, which are associated with $G$-invariant tensors of $\mathfrak{g}$. Given $\mathfrak{g}$, the number of independent (primitive) invariants is precisely the rank of $\mathfrak{g}$. Moreover, some of the variables $\nu_n$ might be identically zero --- simple examples include odd invariants $\nu_{2p+1}$, $p \in \mathbb{N}$, for the algebras $\mathfrak{so}(2N)$, $\mathfrak{so}(2N+1)$, and $\mathfrak{sp}(2N)$. The classification of the independent $\nu_n$ is related to the study of the independent higher-spin currents of the PCM \cite{Evans:1999mj}. For the scope of this paper, we will not worry about model-dependent details and only assume that $N$ in \eqref{scalars_defn} is large enough to have a complete (potentially redundant) set of variables for a given algebra $\mathfrak{g}$.

Having defined these quantities, we now consider the Lagrangian
\begin{align}\label{afsm_defn}
    \mathcal{L}_{\text{AFSM}} = \frac{1}{2} \tr ( j_+ j_- ) + \tr ( v_+ v_- ) + \tr ( j_+ v_- + j_- v_+ ) + E ( \nu_2 , \ldots , \nu_N ) \, ,
\end{align}
where $E$ is a differentiable, but otherwise arbitrary, function of the $N - 1$ independent variables $\nu_n$. We refer to this object $E$ as the interaction function. When $E = 0$, the solution to the algebraic equation of motion for the auxiliary field $v_{\pm}$ is simply
\begin{align}\label{E_zero_soln}
    v_{\pm} \deq - j_{\pm} \, ,
\end{align}
where we have introduced the notation $\deq$ to indicate two quantities which coincide when the auxiliary field equation of motion is satisfied. Substituting the solution (\ref{E_zero_soln}) and the assumption $E = 0$ into the AFSM Lagrangian (\ref{afsm_defn}) then gives
\begin{align}
    \mathcal{L}_{\text{AFSM}} \big\vert_{E = 0} \deq - \frac{1}{2} \tr ( j_+ j_- ) = \mathcal{L}_{\text{PCM}} \, .
\end{align}
In this sense, the AFSM is a deformation of the PCM where the ``deformation parameter'' is the function $E$, since setting this function to zero recovers the undeformed model.

Let us now discuss the $v_{\pm}$ equation of motion when $E \neq 0$, which takes the form
\begin{align}\label{v_eom}
    0 \deq j_{\pm} + v_{\pm} + \sum_{n = 2}^{N} n \frac{\partial E}{\partial \nu_n} \tr ( v_{\pm}^n ) v_{\mp}^{A_1} v_{\mp}^{A_2} \ldots v_{\mp}^{A_{n-1}} T^{A_n} \tr ( T_{(A_1} T_{A_2} \ldots T_{A_{n} )} ) \, ,
\end{align}
where we symmetrize over collections of indices as $X_{(AB)} = \frac{1}{2} \left( X_{AB} + X_{B A} \right)$ and so on. We will encounter equations with structures similar to (\ref{v_eom}) in the generalizations of the AFSM to symmetric and semi-symmetric spaces, with or without WZ term, where the only difference will be that $j_{\pm}$ and $v_{\pm}$ are replaced with their projections onto particular subspaces of $\mathfrak{g}$. It is useful to pause and explain certain implications of equation (\ref{v_eom}), since we may then apply the analogous results in later sections with only minor modifications.

The key ingredient which we need to unravel the consequences of equation (\ref{v_eom}) is a simple identity which applies to the generators $T_A$ of any semi-simple Lie algebra $\mathfrak{g}$ and which is sometimes called the generalized Jacobi identity (e.g. in \cite{vanRitbergen:1998pn}). If $\tensor{f}{_A_B^C}$ are the structure constants of $\mathfrak{g}$, which are defined by the relation
\begin{align}
    [ T_A, T_B ] = \tensor{f}{_A_B^C} T_C \, ,
\end{align}
then any trace of $n$ generators $T_{A_1}$, $\ldots$, $T_{A_n}$ satisfies the equation
\begin{align}\label{generalized_jacobi}
    \sum_{i=1}^{n} \tensor{f}{_C_{A_i}^B} \tr \left( T_{A_1} \ldots T_{A_{i-1}} T_B T_{A_{i+1}} \ldots T_{A_n} \right) = 0 \, ,
\end{align}
where in the $i$-th term of the sum, the corresponding index $A_i$ of the $i$-th generator in the trace is replaced with the index $B$. We refer the reader to Appendix B of \cite{Bielli:2024ach} for details on the derivation of this identity and on the implication which we are about to discuss.\footnote{More generally, see Section 2 and Appendix A of the same work \cite{Bielli:2024ach} for additional information about the higher-spin auxiliary field sigma model which we review here.}

A simple corollary of the generalized Jacobi identity (\ref{generalized_jacobi}) is the statement that, if 
\begin{align}
    M_{A_1 \ldots A_n} = \tr \left( T_{(A_1} \ldots T_{A_n)} \right) \, ,
\end{align}
is a totally symmetric product of generators, then one has
\begin{align}\label{nice_jacobi}
    0 = \tensor{f}{_C_{( A_1}^B} M_{A_{2} \ldots A_{n} ) B} \, .
\end{align}
This result will be useful in analyzing commutators involving one auxiliary field and one Maurer-Cartan form, assuming that the auxiliary field equation of motion (\ref{v_eom}) is satisfied, since such a commutator introduces a factor of structure constants in exactly the way that appears in (\ref{nice_jacobi}). More precisely, we find that
\begin{align}\label{last_term_vanish}
    [ v_{\mp} , j_{\pm} ] &\deq \left[ v_{\mp} , - v_{\pm} - \sum_{n = 2}^{N} n \frac{\partial E}{\partial \nu_n}  \tr ( v_{\pm}^n ) v_{\mp}^{A_1} \ldots v_{\mp}^{A_{n-1}} T^B \tr ( T_{(B} T_{A_1} \ldots T_{A_{n-1} )} ) \right] \nonumber \\
    &= [ v_\pm , v_\mp ] + \sum_{n=2}^{N} n \frac{\partial E}{\partial \nu_n}  \tr ( v_{\pm}^n ) v_{\mp}^{A_1} \ldots v_{\mp}^{A_{n-1}} v_{\mp}^C \tr ( T_{(B} T_{A_1} \ldots T_{A_{n-1} )} ) \tensor{f}{^B_C^D} T_D \nonumber \\
    &= [ v_\pm , v_\mp ] \, ,
\end{align}
where in the first line we have eliminated $j_{\pm}$ in favor of $v_{\pm}$ using (\ref{v_eom}), and in going to the final line we have used the relation (\ref{nice_jacobi}), noting that the required symmetrization over the index $C$ of the structure constants is implemented by the contraction against the totally symmetric tensor $v_{\mp}^{A_1} \ldots v_{\mp}^{A_{n-1}} v_{\mp}^C$. We conclude that
\begin{align}\label{fundamental_commutator}
    [ v_{\mp} , j_{\pm} ] &\deq [ v_\pm , v_\mp ] \, ,
\end{align}
which is the fundamental commutator identity that is needed to establish the classical integrability of any member of this class of models.

We now turn to the discussion of integrability, which concerns the equation of motion for the physical field $g$. By considering a variation $g \to g e^{\epsilon}$ and demanding stationarity of the action associated with the Lagrangian (\ref{afsm_defn}), one finds the Euler-Lagrange equation
\begin{align}\label{j_eom}
    \partial_+ ( j_- + 2 v_- ) + \partial_- ( j_+ + 2 v_+ ) = 2 \left( [ v_- , j_+ ] + [ v_+, j_- ] \right) \, .
\end{align}
We note that (\ref{j_eom}) is the equation of motion for the physical, propagating degree of freedom in this model. In contrast, the auxiliary field equation of motion (\ref{v_eom}) should be viewed differently; one can always use the latter equation to eliminate $v_{\pm}$ in favor of $j_{\pm}$, obtaining a single equation of motion for the group-valued field $g$. Ordinarily, one would say that the resulting model is (weakly) classically integrable if the single equation of motion for $g$ is equivalent to the flatness of a one-form $\mathfrak{L}_{\pm}$ called the Lax connection, for any value of a spectral parameter $z \in \mathbb{C}$, on which $\mathfrak{L}_{\pm}$ is assumed to depend meromorphically. Accordingly, we say that the original auxiliary field model is (weakly) classically integrable if integrating out the auxiliary fields gives rise to a model for $g$ which is (weakly) classically integrable in this ordinary sense. This is equivalent to the statement that there exists a Lax connection $\mathfrak{L}_\pm$, which may depend on both $j_\pm$ and $v_\pm$, whose flatness is equivalent to the $g$-field equation of motion when one assumes that the auxiliary field Euler-Lagrange equation is satisfied. This motivates us to consider the structure of equation (\ref{j_eom}) when (\ref{v_eom}) holds; in particular, since the latter implies the commutator identity (\ref{fundamental_commutator}), the $g$-field equation of motion can be written as
\begin{align}\label{j_eom_deq}
    \partial_+ ( j_- + 2 v_- ) + \partial_- ( j_+ + 2 v_+ ) \deq 0 \, ,
\end{align}
when the auxiliary field equation of motion is obeyed. This suggests that one should define a combination of the Maurer-Cartan form and auxiliary field,
\begin{align}
    \mathfrak{J}_{\pm} = - \left( j_{\pm} + 2 v_{\pm} \right) \, ,
\end{align}
with the property that $\mathfrak{J}_{\pm}$ reduces to $j_{\pm}$ in the undeformed theory with $E = 0$ and $v_{\pm} = - j_{\pm}$, and such that the $g$-field Euler-Lagrange equation can be written as the conservation of $\mathfrak{J}_{\pm}$ when the auxiliary field equation of motion is satisfied:
\begin{align}\label{dot_g_eom}
    \partial_+ \mathfrak{J}_- + \partial_- \mathfrak{J}_+ \deq 0 \, .
\end{align}
A zero-curvature representation for (\ref{dot_g_eom}) is furnished by the Lax connection
\begin{align}\label{afsm_lax}
    \mathfrak{L}_{\pm} = \frac{j_{\pm} \pm z \mathfrak{J}_{\pm}}{1 - z^2} \, ,
\end{align}
since using the fundamental commutator relation (\ref{fundamental_commutator}), one finds that
\begin{align}\label{afsm_lax_curv}
    \partial_+ \mathfrak{L}_- - \partial_- \mathfrak{L}_+ + [ \mathfrak{L}_+ , \mathfrak{L}_- ] \deq \frac{1}{1 - z^2} \left( \partial_+ j_- - \partial_- j_+ + [ j_+ , j_- ] - z \left( \partial_+ \mathfrak{J}_- + \partial_- \mathfrak{J}_+ \right) \right) \, .
\end{align}
The first three terms in the parentheses of the right side of (\ref{afsm_lax_curv}) vanish due to the Maurer-Cartan identity (\ref{mc_identity}), which means that $\partial_+ \mathfrak{L}_- - \partial_- \mathfrak{L}_+ + [ \mathfrak{L}_+ , \mathfrak{L}_- ] \deq 0$ if and only if $\partial_+ \mathfrak{J}_- + \partial_- \mathfrak{J}_+ \deq 0$, establishing the Lax representation for (\ref{dot_g_eom}). Finally, let us note that the conserved charges constructed from the monodromy matrix of the Lax (\ref{afsm_lax}) can be shown to be in involution, which was demonstrated in \cite{Bielli:2024ach} using the Maillet $r$/$s$ formalism \cite{MAILLET198654,MAILLET1986401}.

\subsection{AFSM with Wess-Zumino Term}\label{sec:AFSM_WZ}

A well-known integrable \cite{ABDALLA1982181} deformation of the principal chiral model is the addition of a Wess-Zumino term \cite{WESS197195,Novikov:1982ei}. For a particular choice of the relative coefficient between the WZ term and the PCM kinetic term, this deformation defines the Wess-Zumino-Witten (WZW) model \cite{Witten:1983ar}, which gives rise to a conformal field theory at the quantum level. In contrast, the PCM with WZ-term (or PCM-WZ for short) at a generic value of this relative coefficient is merely a classically conformal field theory.

It has been shown that the PCM-WZ can be combined with the auxiliary field deformations reviewed in Section \ref{sec:afsm_review} to obtain a doubly-deformed family of integrable models. This was shown first for interaction functions $E ( \nu_2 )$ of a single variable in \cite{Fukushima:2024nxm} from the perspective of $4d$ Chern-Simons theory, and then for general interaction functions  $E ( \nu_2 , \ldots, \nu_N )$ in \cite{Bielli:2024ach}. We now review a few salient aspects of this latter, more general construction.

First we will restore a general constant $\hay$ which multiplies the Lagrangian (\ref{pcm_lc}), in addition to a second coefficient $\kay$ of the Wess-Zumino term, and write the Lagrangian of the undeformed PCM-WZ as
\begin{align}\label{PCM_WZ}
    S_{\text{PCM-WZ}} = - \frac{\hay}{2} \int_{\Sigma} d^2 \sigma \, \tr ( j_+ j_- ) + \frac{\kay}{6} \int_{\mathcal{M}_3} d^3 x \, \epsilon^{ijk} \tr \left( j_i [ j_j, j_k ] \right) \, .
\end{align}
The second term in (\ref{PCM_WZ}) is an integral over a three-dimensional manifold $\mathcal{M}_3$ whose boundary is the $2d$ worldsheet $\Sigma$. We use lowercase middle Latin letters like $i$, $j$, $k$ for indices on $\mathcal{M}_3$, to distinguish them from the early lowercase Greek letters $\alpha$, $\beta$, etc. which indicate indices on $\Sigma = \partial \mathcal{M}_3$. When $\hay = 1$ and $\kay = 0$, (\ref{PCM_WZ}) reduces to the action of the usual PCM; more generally, the physics of the model is controlled by the parameter $\frac{\kay}{\hay}$.

To promote (\ref{PCM_WZ}) to the analogous theory deformed by auxiliary fields, which we refer to as the AFSM-WZ, we proceed by replacing the integrand in the $2d$ part of the action with the standard AFSM Lagrangian (\ref{afsm_defn}) and leaving the Wess-Zumino term unmodified:
\begin{align}\label{family_with_WZ}
    S_{\text{AFSM-WZ}} &= \hay \int_{\Sigma} d^2 \sigma \, \left( \frac{1}{2} \mathrm{tr} ( j_+ j_- ) + \mathrm{tr} ( v_+ v_- ) + \mathrm{tr} ( j_+ v_- + j_- v_+ ) + E ( \nu_2 , \ldots , \nu_N )  \right) \nonumber \\
    &\qquad + \frac{\kay}{6} \int_{\mathcal{M}_3} d^3 x \, \epsilon^{ijk} \tr \left( j_i [ j_j, j_k ] \right) \, .
\end{align}
Because the auxiliary field $v_{\pm}$ does not appear anywhere in the Wess-Zumino term of (\ref{family_with_WZ}), its equation of motion is identical to that of the usual AFSM, equation (\ref{v_eom}). By the same reasoning as in Section \ref{sec:afsm_review}, it follows that the fundamental commutator relation (\ref{fundamental_commutator}) also holds in the AFSM-WZ when the auxiliary field Euler-Lagrange equation is satisfied.

In contrast, the equation of motion for $g$ is modified by the presence of the Wess-Zumino term. Under a variation $g \to g e^{\epsilon}$, the variation of the $2d$ part of the action is the same as that of the AFSM, but one also generates a contribution
\begin{align}\label{final_delta_SWZ}
    \delta S_{\text{WZ}} = \frac{1}{2} \int_{\mathcal{M}_3} d^3 x \, \epsilon^{ijk} \partial_i \tr \left( \epsilon [ j_j , j_k ] \right) \, ,
\end{align}
which is identical to the corresponding variation in the PCM-WZ (since, again, the WZ term is independent of auxiliaries). The integrand in the variation (\ref{final_delta_SWZ}) is manifestly a total derivative and thus localizes to an integral over the boundary $\Sigma$, as required in order to have a well-defined $2d$ theory.\footnote{We refer the reader to Section 2.2 of the review \cite{Hoare:2021dix}, or Section 5.1 and Appendix A.3 of \cite{Bielli:2024ach}, for further details on these manipulations and related discussions.}

Combining the contributions from the variations of the $2d$ integral and the WZ term, the full equation of motion for $g$ in the AFSM-WZ is
\begin{align}\label{AFSM_WZ_g_eom}
    \left( \hay - \kay \right) \partial_+ j_- + \left( \hay + \kay \right) \partial_- j_+ + 2 \hay \left( \partial_+ v_- + \partial_- v_+ + [ j_+ , v_- ] + [ j_- ,v_+ ] \right) = 0 \, .
\end{align}
Precisely as before, due to the fundamental commutator relation (\ref{fundamental_commutator}), when the auxiliary field equation of motion is satisfied this can be written as
\begin{align}\label{PCM_WZ_eom_dotted}
    \partial_+ \left( \hay \mathfrak{J}_- + \kay j_- \right) + \partial_- \left( \hay \mathfrak{J}_+ - \kay j_+ \right) \deq 0 \, ,
\end{align}
where we have defined $\mathfrak{J}_{\pm} = - \left( j_\pm + 2 v_{\pm} \right)$ exactly as in the AFSM without WZ term.

The AFSM-WZ is also (weakly) classically integrable, in the same sense that we have described above for the AFSM. To see this, we define the Lax connection
\begin{align}\label{AFSM_WZ_lax}
    \mathfrak{L}_{\pm} = \frac{ \left( j_{\pm} \mp \frac{\kay}{\hay} \mathfrak{J}_{\pm} \right) \pm z \left( \mathfrak{J}_{\pm} \mp \frac{\kay}{\hay} j_{\pm} \right) }{1 - z^2} \, .
\end{align}
Using the commutator identity (\ref{fundamental_commutator}), one finds
\begin{align}\label{lax_curv_afsm_wz}
    \partial_+ \mathfrak{L}_- - \partial_- \mathfrak{L}_+ + [ \mathfrak{L}_+ , \mathfrak{L}_- ] \deq \frac{1}{1 - z^2} &\Bigg( \partial_+ j_- - \partial_- j_+ + [ j_+ , j_- ]  \\
    & \, + \left( \frac{\kay}{\hay} - z \right) \left( \partial_+ \left( \mathfrak{J}_- + \frac{\kay}{\hay} j_- \right) + \partial_- \left(  \mathfrak{J}_+ - \frac{\kay}{\hay} j_+ \right) \right) \Bigg) \, .
    \nonumber
\end{align}
Much like in the argument of Section \ref{sec:afsm_review}, the first line on the right side of (\ref{lax_curv_afsm_wz}) is identically zero due to the Maurer-Cartan identity for $j_{\pm}$. The remaining terms on the second line then vanish if and only if the equation of motion (\ref{PCM_WZ_eom_dotted}) is satisfied, which means that (\ref{AFSM_WZ_lax}) gives the desired Lax representation for the equations of motion.

\section{Symmetric Space Sigma Models}\label{sec:SSSM}

The auxiliary field machinery reviewed in Section \ref{sec:review} gives a powerful toolkit for constructing integrable deformations of sigma models whose target space is a Lie group $G$. However, of course not all target spaces of interest are group manifolds. Even very simple (and physically relevant) symmetric space examples such as $n$-dimensional spheres
\begin{align}
    S^n \cong \frac{SO ( n + 1 ) }{SO ( n ) } \, ,
\end{align}
and $\mathrm{AdS}$ spacetimes,
\begin{align}
    \mathrm{AdS}_n \cong \frac{ SO ( 2 , n - 1 ) }{SO ( 1, n-1 ) } \, ,
\end{align}
are realized as cosets and are not group manifolds (except for special cases like $S^3$). 

It is natural to ask whether auxiliary field deformations can also be applied to sigma models with more general target spaces of this form. In this section, we will answer this question in the affirmative. We show that one can activate general, higher-spin auxiliary field couplings for sigma models on symmetric homogeneous spaces (which we also refer to as symmetric cosets) while preserving classical integrability via a mechanism which is very similar to that which we have seen for the AFSM.

We begin by reviewing some standard features of sigma models whose target spaces are symmetric cosets, which are often referred to as symmetric space sigma models.

\subsection{Generalities on SSSM}\label{sec:sssm_review}

Consider a field $g: \Sigma \to G / H$ valued in the coset of a Lie group $G$ by a subgroup $H$. Let $\mathfrak{g}$ be the Lie algebra of $G$ and let $\mathfrak{h}$ be the Lie algebra of $H$. We will also write $\mathfrak{g}_0$ for $\mathfrak{h}$, and we define $\mathfrak{g}_2$ as the orthogonal complement so that one has a decomposition
\begin{align}
    \mathfrak{g} = \mathfrak{g}_0 \oplus \mathfrak{g}_2 \, .
\end{align}
In order to have a symmetric space, we will assume that
\begin{align}\label{symm_cond}
    [ \mathfrak{g}_{n} , \mathfrak{g}_m ] \subset \mathfrak{g}_{( n + m ) \, \text{mod } 4 } \, .
\end{align}
All Lie algebra valued fields can then be decomposed according to their projections onto $\mathfrak{g}_0$ and $\mathfrak{g}_2$, which we will indicate with superscripts. For instance, the left-invariant Maurer-Cartan form $j_\alpha = g^{-1} \partial_\alpha g$ enjoys the decomposition
\begin{align}
    j_\alpha = j_\alpha^{(0)} + j_\alpha^{(2)} \, , \qquad j_\alpha^{(n)} \in \mathfrak{g}_n \, .
\end{align}
It may be helpful to comment on the physical interpretations of the two components $j_\alpha^{(0)}$ and $j_\alpha^{(2)}$, which are rather different. Since we are interested in the coset $G / H$ of $G$ by $H$ on the right, roughly speaking, we would like to consider the action of right-multiplying by an element of $H$ as ``pure gauge'' and thus we might expect that $j_\pm^{(0)} \in \mathfrak{g}_0 = \mathfrak{h}$ admits an interpretation as a sort of gauge field. Adopting this perspective, one could therefore define natural notions of covariant derivative,
\begin{align}\label{SSSM_covariant_deriv}
    D_\alpha = \partial_\alpha + \left[ j_\alpha^{(0)} \; , \; \cdot \, \right] \, ,
\end{align}
and field strength,
\begin{align}\label{SSSM_field_strength}
    F_{\alpha \beta}^{(0)} = \partial_\alpha j_\beta^{(0)} - \partial_\beta j_\alpha^{(0)} + [ j_\alpha^{(0)} , j_\beta^{(0)} ] \, ,
\end{align}
associated with the ``gauge field'' $j_\alpha^{(0)}$. On the other hand, under right-multiplication $g \to g' = g h$ by an arbitrary element $h \in H$, one finds that $j_\alpha^{(2)}$ transforms adjointly. This suggests that we interpret $j_\alpha^{(2)}$ as the primary physical degree of freedom in this model, which is charged under gauge transformations of the background gauge field $j_\alpha^{(0)}$. From this perspective, it is natural to propose a Lagrangian which resembles that of the PCM (\ref{pcm_lc}) but which involves the projection $j_\alpha^{(2)}$:
\begin{align}\label{SSSM}
    \mathcal{L}_{\text{SSSM}} = \frac{1}{2} \eta^{\alpha \beta} \tr \left( j_\alpha^{(2)} j_\beta^{(2)} \right) = - \frac{1}{2} \tr \left( j_+^{(2)} j_-^{(2)} \right) \, .
\end{align}
Equation (\ref{SSSM}) defines the undeformed symmetric space sigma model on the coset $G / H$. Although $j_\alpha^{(0)}$ does not appear explicitly in the Lagrangian, it is involved in the dynamics implicitly, since a variation $g \to g e^{\epsilon}$ of the full group-valued field $g$ affects both $j_\alpha^{(0)}$ and $j_\alpha^{(2)}$. Demanding stationarity of the action under such a variation gives the equation of motion
\begin{align}\label{sssm_eom}
    D^\alpha j_\alpha^{(2)} = 0 \, ,
\end{align}
where the dependence on $j_\alpha^{(0)}$ is hidden in the action of the covariant derivative $D_\alpha$.

In order to demonstrate the classical integrability of the symmetric coset model, one can show that the equation of motion (\ref{sssm_eom}) is equivalent to the flatness of the Lax connection
\begin{align}\label{sssm_lax}
    \mathfrak{L}_{\pm} = j_{\pm}^{(0)} + \frac{z \mp 1}{z \pm 1} j_{\pm}^{(2)} \, ,
\end{align}
at any value of the spectral parameter $z \in \mathbb{C}$.

For further details regarding the SSSM, such as the derivation of the equation of motion (\ref{sssm_eom}) and the proof of its equivalence to the flatness of the Lax connection (\ref{sssm_lax}), see Section 4 of \cite{Zarembo:2017muf}, Section 1.3 of \cite{yoshida2021yang}, or Section 2.2.3 of \cite{Seibold:2020ouf}.

\subsection{SSSM with Auxiliary Fields}

Next let us see how the symmetric coset model reviewed in Section \ref{sec:sssm_review} can be deformed by including interactions via auxiliary fields. We will sometimes refer to the resulting deformed theory as the auxiliary field symmetric space sigma model, or AF-SSSM. 

We again introduce a Lie algebra valued auxiliary field $v_\alpha$ and decompose it as
\begin{align}\label{SSSM_aux_decomp}
    v_\alpha = v_\alpha^{(0)} + v_\alpha^{(2)} \, ,
\end{align}
where $v_\alpha^{(n)} \in \mathfrak{g}_n$, just as in the decomposition of $j_\alpha$.

In the undeformed SSSM, we have argued that one should interpret $j_\alpha^{(0)}$ as a sort of background gauge field, while in contrast we think of $j_\alpha^{(2)}$ as the main field of interest. Building upon this intuition, we will introduce interactions only between $j_\alpha^{(2)}$ and the corresponding projection $v_\alpha^{(2)} \in \mathfrak{g}_2$. To do this, we start by defining a collection of scalars $\nu_n$,
\begin{align}
    \nu_n = \tr \Big( \underbrace{ v^{(2)}_+ \ldots v^{(2)}_+}_{n \, \text{ times}} \Big) \tr \Big( \underbrace{ v^{(2)}_- \ldots v^{(2)}_-}_{n \, \text{ times}} \Big)  \, ,
\end{align}
by analogy with the definition (\ref{scalars_defn}) in the case of the AFSM. 
The only difference in the present setting is that we use $v_\alpha^{(2)}$ rather than $v_\alpha$. Continuing to follow the procedure which we carried out in Section \ref{sec:afsm_review}, 
we then consider the Lagrangian
\begin{align}\label{SSSM_deformed}
    \mathcal{L} = \frac{1}{2} \tr ( j_+^{(2)} j_-^{(2)} ) + \tr ( v_+^{(2)} v_-^{(2)} ) + \tr ( j_+^{(2)} v_-^{(2)} + j_-^{(2)} v_+^{(2)} ) + E \left( \nu_2 , \ldots , \nu_N \right) \, .
\end{align}
Note that, analogously to what we mentioned in subsection \ref{sec:afsm_review}, depending on the details of $\mathfrak{g}$ -- and now also $\mathfrak{g}_0$ and $\mathfrak{g}_2$ -- some of the variables $\nu_n$ might be zero or dependent upon one another. In particular, since we are now constructing $\nu_n$ in terms of invariant tensors restricted to $\mathfrak{g}_2$, the independent ones should be associated with $H$-invariant tensors. 
A description of the primitive invariants restricted to symmetric spaces $G/H$ can be found in \cite{Evans:2000qx}; see also \cite{Evans:2001sz} for the cases including exceptional finite-dimensional Lie algebras. In these papers, a detailed classification of independent classical higher-spin currents for symmetric space sigma models was performed. Interesting examples to mention from the analysis of \cite{Evans:2000qx} are the symmetric spaces $SU(N)/SO(N)$ and $SU(2N)/Sp(2N)$, which possess both odd and even spin higher-spin currents, and hence in our case $\nu_n$ with both $n=2p+1$ and $n=2p$. All other symmetric spaces studied in \cite{Evans:2000qx}, specifically the spaces $SU(p+q)/S(U(p)\times U(q))$, $SO(p+q)/SO(p)\times SO(q)$, $SO(2N)/SO(N)\times SO(N)$,  $Sp(2p+2q)/Sp(2p)\times Sp(2q)$, $SO(2N)/U(N)$,  and $Sp(2N)/U(N)$, possess only even spin currents, meaning that in our case only the $\nu_{n}$ with $n=2p$, $p \in \mathbb{N}$, are non-trivial for these examples. As in the case of the PCM, for the scope of this paper, we will not worry about model-dependent specific details and only assume that $N$ in \eqref{scalars_defn} is large enough to have a complete (potentially redundant) set of variables for a given algebra $\mathfrak{g}$.

We now come back to (\ref{SSSM_deformed}), compared to the PCM case.
There are again two equations of motion arising from (\ref{SSSM_deformed}), one for the auxiliary field $v_\alpha$ and one associated with the physical field $g$. The derivation of these equations of motion is carried out in Appendix \ref{app:SSSM_eom}; here we will simply quote and interpret the results.

First, the auxiliary field equation of motion is
\begin{align}\label{SSSM_auxiliary_eom}
    0 &\deq j_{\pm}^{(2)} + v_{\pm}^{(2)} \\
    &\quad + \sum_{n = 2}^{N} n \frac{\partial E}{\partial \nu_n} \tr \left( \Big( v_{\pm}^{(2)} \Big)^n \right) \Big( v_{\mp}^{(2)} \Big)^{A_1} \left( v_{\mp}^{(2)} \right)^{A_2} \ldots \left( v_{\mp}^{(2)} \right)^{A_{n-1}} T^{A_n} \tr ( T_{(A_1} T_{A_2} \ldots T_{A_{n} )} ) \, .
    \notag
\end{align}
Two comments are in order. First, in (\ref{SSSM_auxiliary_eom}) we have written quantities like $\left( v_{\pm}^{(2)} \right)^A$, which are the expansion coefficients of $v_{\pm}^{(2)}$ in the basis $T_A$ of generators of the full Lie algebra $\mathfrak{g}$. That is, one has the expansion
\begin{align}\label{projected_expansion}
    v_{\pm}^{(2)} = \big( v_{\pm}^{(2)} \big)^A T_A \, .
\end{align}
Since $\mathfrak{g}_2$ is a subspace of $\mathfrak{g}$, any element of $\mathfrak{g}_2$ also admits an expansion in the generators $T_A$ of the full Lie algebra $\mathfrak{g}$, even though many of the expansion coefficients may be vanishing if the corresponding generators lie within the orthogonal complement of $\mathfrak{g}_2$. Therefore the expansion (\ref{projected_expansion}) is well-defined. Alternatively, one could have taken a different approach and chosen a set of generators $T^{(2)}_a$ for the Lie algebra $\mathfrak{g}_2$, and performed an expansion
\begin{align}\label{projected_expansion_different}
    v_{\pm}^{(2)} = \big( v_{\pm}^{(2)} \big)^a T_a^{(2)} \, ,
\end{align}
where the lowercase early Latin index $a$ labels the generators $T_a^{(2)}$ of $\mathfrak{g}_2$ (these generators need not be a subset of the $T_A$, in general; one is only guaranteed that the span of the $T_a^{(2)}$ is a subspace of the span of the $T_A$). However, for the argument which we are interested in, it will not be necessary to refer to the generators $T_a^{(2)}$ directly, so we will always work directly with expansions in the generators $T_A$ of $\mathfrak{g}$ as in equations (\ref{SSSM_auxiliary_eom}) and (\ref{projected_expansion}).

The second comment is that the $v_{\pm}$ Euler-Lagrange equation (\ref{SSSM_auxiliary_eom}) takes exactly the same form as the corresponding equation of motion (\ref{v_eom}) for the AFSM, with the only difference being the replacements $j_\alpha \to j_\alpha^{(2)}$ and $v_\alpha \to v_\alpha^{(2)}$. As a result, we can apply the same argument involving the generalized Jacobi identity, which was reviewed in Section \ref{sec:afsm_review}, almost verbatim in the present context. Doing this yields the AF-SSSM version of the fundamental commutator identity (\ref{fundamental_commutator}),
\begin{align}\label{sssm_fundamental_commutator}
    [ v_{\mp}^{(2)} , j_{\pm}^{(2)} ] \deq [ v_{\pm}^{(2)} , v_{\mp}^{(2)} ] \, .
\end{align}
The relation (\ref{sssm_fundamental_commutator}) will play a starring role in the proof of classical integrability for the AF-SSSM, much as the analogous identity (\ref{fundamental_commutator}) did for the AFSM.

The equation of motion for the field $g$, which is also derived in Appendix \ref{app:SSSM_eom}, is
\begin{align}\label{SSSM_eom_unsimplified}
    D_- \left( j_+^{(2)} + 2 v_+^{(2)} \right) + D_+ \left( j_-^{(2)} + 2 v_-^{(2)} \right) = 2 \left(   [ v_-^{(2)} , j_+^{(2)} ] + [ v_+^{(2)} , j_-^{(2)} ] \right) \, ,
\end{align}
which takes almost the same form as for the deformed PCM, up to promoting partial derivatives $\partial_{\pm}$ to covariant derivatives $D_{\pm}$, and appending ${}^{(2)}$ superscripts to all fields. We also note that the left side of (\ref{SSSM_eom_unsimplified}) belongs to $\mathfrak{g}_2$ while the right side belongs to $\mathfrak{g}_0$, so in fact each side of this equation must vanish separately. However, the vanishing of the right side of (\ref{SSSM_eom_unsimplified}) is strictly weaker than the commutator identity (\ref{sssm_fundamental_commutator}), which is itself weaker than the full auxiliary field equation of motion (\ref{sssm_fundamental_commutator}). 

Regardless, for our purposes it is sufficient that when the auxiliary field equation of motion is satisfied, the right side of (\ref{SSSM_eom_unsimplified}) is equal to zero, and the only remaining condition for the $g$-field Euler-Lagrange equation (\ref{SSSM_eom_unsimplified}) is the vanishing of the left side:
\begin{align}\label{SSSM_eom_simplified}
    D_- \mathfrak{J}_+^{(2)} + D_+ \mathfrak{J}_-^{(2)} \deq 0 \, ,
\end{align}
where as before we have defined the combination $\mathfrak{J}_{\pm} = - \left( j_{\pm} + 2 v_{\pm} \right)$ and its projection
\begin{align}
    \mathfrak{J}_{\pm}^{(2)} = - \left( j_{\pm}^{(2)} + 2 v_{\pm}^{(2)} \right) \, .
\end{align}
Finally, let us also briefly discuss the decomposition of the Maurer-Cartan identity (\ref{mc_identity}). In terms of the field strength (\ref{SSSM_field_strength}), this condition can be written as
\begin{align}
    F_{\alpha \beta}^{(0)} + D_\alpha j_\beta^{(2)} - D_\beta j_\alpha^{(2)} + [ j_\alpha^{(2)} , j_\beta^{(2)} ] = 0 \, .
\end{align}
As we mentioned, for a symmetric space, the projection of any equation onto the two subspaces $\mathfrak{g}_0$ and $\mathfrak{g}_2$ must separately hold. Note that $F_{\alpha \beta} \in \mathfrak{g}_0$, $[ j_\alpha , j_\beta ] \in \mathfrak{g}_0$, while $D_\alpha j_\beta^{(2)} \in \mathfrak{g}_2$. So in fact the Maurer-Cartan identity gives us two independent conditions,
\begin{align}\label{SSSM_two_maurer_cartan}
    F_{\alpha \beta}^{(0)} + [ j_\alpha^{(2)} , j_\beta^{(2)} ] = 0 \, , \qquad D_\alpha j_\beta^{(2)} - D_\beta j_\alpha^{(2)}  = 0 \, .
\end{align}

\subsection{Classical Integrability}\label{sec:afsssm_integrability}

We now address the integrability of the AF-SSSM constructed above, and in particular the zero-curvature representation for the $g$-field equation of motion (\ref{SSSM_eom_simplified}).

We begin by proposing the Lax connection for the SSSM deformed by auxiliary fields,
\begin{align}\label{SSSM_lax_ansatz_later}
    \mathfrak{L}_{\pm} = j_\pm^{(0)} + \frac{ \left( z^2 + 1 \right) j_{\pm}^{(2)} \mp 2 z \mathfrak{J}_{\pm}^{(2)}}{z^2 - 1} \, .
\end{align}
As a check, let us consider the undeformed limit, where
\begin{align}
    E ( \nu_2 , \ldots , \nu_N ) = 0 \, , \qquad \mathfrak{J}_{\pm} \deq  j_{\pm} \, .
\end{align}
In this case, the components of (\ref{SSSM_lax_ansatz_later}) reduce to 
\begin{align}
    \mathfrak{L}_+ &\deq j_+^{(0)} + \frac{z^2 - 2 z + 1}{z^2 - 1} j^{(2)}_+ = j_+^{(0)} + \frac{z-1}{z+1} j^{(2)}_+ \, ,
\end{align}
and
\begin{align}
    \mathfrak{L}_- &\deq j_-^{(0)} + \frac{z^2 + 2 z + 1}{z^2 - 1} j^{(2)}_- = j_-^{(0)} + \frac{z + 1}{z - 1} j^{(2)}_- \, ,
\end{align}
which agree with the expression for the Lax connection (\ref{sssm_lax}) of the undeformed SSSM.

Next we consider the deformed case, including the coupling to auxiliary fields. We are interested in computing the curvature of the Lax connection (\ref{SSSM_lax_ansatz_later}), which we can write as
\begin{align}
    d_{\mathfrak{L}} \mathfrak{L} = \partial_+ \mathfrak{L}_- - \partial_- \mathfrak{L}_+ + [ \mathfrak{L}_+ , \mathfrak{L}_- ] \, ,
\end{align}
where we have defined the covariant exterior derivative with respect to the connection $\mathfrak{L}$ as
\begin{align}
    d_{\mathfrak{L}} = d + \mathfrak{L} \, \wedge \, \, .
\end{align}
We have collected the details of this computation in Appendix \ref{app:SSSM_lax}. Here we will only point out that the main ingredient in these manipulations is the fundamental commutator identity (\ref{sssm_fundamental_commutator}) for the AF-SSSM, which implies useful relations like
\begin{align}\label{afsssm_derived_commutators}
    [ \mathfrak{J}_+^{(2)} , \mathfrak{J}_-^{(2)} ] \deq [ j_+^{(2)} , j_-^{(2)} ] \, , \qquad [ \mathfrak{J}_+^{(2)} , j_-^{(2)} ] \deq [ j_+^{(2)} , \mathfrak{J}_-^{(2)} ] \, ,
\end{align}
as we show in equations (\ref{justify_one}) - (\ref{justify_two}).\footnote{Identical relations hold in the analysis of the SSSM deformed by a combination of $\TT$ and root-$\TT$, as first shown in \cite{Borsato:2022tmu}. This is to be expected since any stress tensor deformation of the SSSM corresponds to the AF-SSSM with a particular choice of interaction function $E ( \nu_2 )$ which depends on only one variable.} Using results of this form, one eventually finds
\begin{align}\label{final_AFSSSM_lax}
    d_{\mathfrak{L}} \mathfrak{L} &\deq F_{+-}^{(0)} + [ j_+^{(2)} , j_-^{(2)} ] + \frac{z^2 + 1}{z^2 - 1} \left( D_+ j_-^{(2)} - D_- j_+^{(2)} \right) + \frac{2 z}{z^2 - 1} \left( D_+ \mathfrak{J}_-^{(2)} + D_- \mathfrak{J}_+^{(2)} \right) \, .
\end{align}
The first two terms in $d_{\mathfrak{L}} \mathfrak{L}$ vanish by the first condition in the Maurer-Cartan identity (\ref{SSSM_two_maurer_cartan}). The third term, proportional to $D_+ j_-^{(2)} - D_- j_+^{(2)}$, vanishes by the second condition in (\ref{SSSM_two_maurer_cartan}). Therefore, the condition $d_{\mathfrak{L}} \mathfrak{L} \deq 0$ for any $z$ is equivalent to the condition
\begin{align}
    D_+ \mathfrak{J}_-^{(2)} + D_- \mathfrak{J}_+^{(2)} \deq 0 \, ,
\end{align}
which is the equation of motion for the model. This proves that every member of our family (\ref{SSSM_deformed}) of deformed symmetric space sigma models is weakly classically integrable, with a Lax connection given by (\ref{SSSM_lax_ansatz_later}).

\section{Semi-Symmetric Space Sigma Model with WZ Term}\label{sec:sSSSM}

Thus far, all of the results concerning auxiliary field deformations in the literature -- including those in the preceding sections of this work -- concern sigma models whose target spaces have only bosonic coordinates. However, sigma models with manifest target-space supersymmetry, such as those describing string propagation on a supergroup or cosets thereof, are also of considerable interest. In some sense, these supersymmetric scenarios are the \emph{most} interesting cases, since they include the sigma models describing Green-Schwarz superstring propagation on supercosets, which was the original setting for the Metsaev-Tseytlin construction \cite{Metsaev:1998it,Bena:2003wd} which initiated applications of integrability to worldsheet string theory.\footnote{The first observation connecting the Green-Schwarz string to supercosets was made earlier, by Henneaux and Merzincescu, in the context of flat space strings \cite{Henneaux:1984mh}.}

For instance, the $\mathrm{AdS}_5 \times S^5$ superstring is described using a sigma model whose target space is the supercoset
\begin{align}
    \frac{PSU ( 2 , 2 \, \vert \, 4 )} { SO ( 1, 4 ) \times SO ( 5 ) } \, ,
\end{align}
and models of string propagation on $\mathrm{AdS}_3 \times S^3 \times T^4$ are based on the supercoset
\begin{align}
    \frac{PSU ( 1, 1 \, \vert \, 2 ) \times PSU ( 1, 1 \, \vert \,  2 ) }{ SU ( 1, 1 ) \times SU ( 2 ) } \, .
\end{align}
Motivated by the importance of these supercoset models, it is natural to wonder whether auxiliary field deformations such as those in the AF-SSSM can be extended to supercosets.

Furthermore, we note that all of the cases that we have considered so far -- including the PCM with WZ term and the symmetric space sigma model -- can be viewed as special cases of a semi-symmetric space sigma model with Wess-Zumino term. For instance, the symmetric space sigma model follows from the semi-symmetric space sigma model with WZ term (sSSSM-WZ) by setting both the WZ term and fermionic part to zero, which corresponds to choosing the Lie supergroup $G$ to be an ordinary Lie group. Likewise, the PCM with WZ term follows from the sSSSM-WZ by choosing a bosonic symmetric space coset of the form $(G \times G) / G$. Therefore, if we prove weak integrability of an auxiliary field deformation of this most general model, all of the other results follow as special cases.

In this section we will show (weak) classical integrability for auxiliary field deformations of the sSSSM with WZ term by giving a Lax representation for the equations of motion.

\subsection{Generalities on sSSSM}

We now assume that $G$ is a Lie supergroup with Lie superalgebra $\mathfrak{g}$. Suppose that the superalgebra is equipped with a $\mathbb{Z}_4$ automorphism $\Omega$, with the property that
\begin{align}
    \Omega^2 = \left( - 1 \right)^F \, , 
\end{align}
and where $( - 1 )^F$ is the fermion number operator which gives the usual $\mathbb{Z}_2$ grading of the superalgebra. One can then perform a decomposition of the Lie superalgebra $\mathfrak{g}$ into subspaces $\mathfrak{g}_n$ with fixed eigenvalues under the action of $\Omega$, namely
\begin{align}
    \Omega \mathfrak{g}_n = i^n \mathfrak{g}_n \, ,
\end{align}
and write
\begin{align}
    \mathfrak{g} = \mathfrak{g}_0 \oplus \mathfrak{g}_1 \oplus \mathfrak{g}_2 \oplus \mathfrak{g}_3 \, .
\end{align}
The subspaces $\mathfrak{g}_n$ for even $n$ contain bosonic generators and the subspaces with odd $n$ correspond to fermionic generators. We also assume that
\begin{align}
    [ \mathfrak{g}_n, \mathfrak{g}_m ] \subset \mathfrak{g}_{( n + m ) \text{ mod } 4} \, .
\end{align}
We define $\mathfrak{h} = \mathfrak{g}_0$ and let $H$ be the subgroup associated with $\mathfrak{h}$. As before, we indicate decompositions of Lie algebra valued quantities with superscripts, such as
\begin{align}\label{four_subspace_decomp}
    j_\alpha = j_\alpha^{(0)} + j_\alpha^{(1)} + j_\alpha^{(2)} + j_\alpha^{(3)} \, , \qquad j_\alpha^{(n)} \in \mathfrak{g}_n \, .
\end{align}
We refer to a sigma model with Lagrangian
\begin{align}\label{sSSSM}
    \mathcal{L}_{\text{sSSSM}} = \frac{1}{2} \mathrm{str} \left( g^{\alpha \beta} j_\alpha^{(2)} j_\beta^{(2)} + \epsilon^{\alpha \beta} j_\alpha^{(1)} j_\beta^{(3)} \right) \, ,
\end{align}
where $\mathrm{str}$ is the supertrace, which is based on a Lie supergroup with the above structure, as a semi-symmetric space sigma model or $\mathbb{Z}_4$ coset model. To ease notation, it will sometimes be convenient to refer to the second (topological) term as $x_0$, i.e. to define
\begin{align}\label{x0_defn}
    x_0 &= \frac{1}{2} \epsilon^{\alpha \beta} \mathrm{str} ( j_\alpha^{(1)} j_\beta^{(3)} ) =  \frac{1}{4} \mathrm{str} \left( j_+^{(1)} j_-^{(3)} - j_-^{(1)} j_+^{(3)} \right) \, .
\end{align}
The equations of motion of (\ref{sSSSM}) are equivalent to the flatness of the Lax connection
\begin{align}
    \mathfrak{L}_{\pm} = j_{\pm}^{(0)} + \sqrt{ \frac{z + 1}{z - 1} } j_{\pm}^{(1)} + \frac{z \mp 1}{z \pm 1} j_{\pm}^{(2)} + \sqrt{ \frac{z - 1}{z+1} } j_{\pm}^{(3)} \, , 
\end{align}
for any value of the spectral parameter $z \in \mathbb{C}$.

For later use, it will be helpful to record the decomposition of the Maurer-Cartan identity along the four subspaces $\mathfrak{g}_n$. One has
\begin{align}
    \partial_+ j_- - \partial_- j_+ + [ j_+ , j_- ] &= \partial_+ \left( j_-^{(0)} + j_-^{(1)} + j_-^{(2)} + j_-^{(3)} \right) - \partial_- \left( j_+^{(0)} + j_+^{(1)} + j_+^{(2)} + j_+^{(3)} \right) \nonumber \\
    &\qquad + [ j_+^{(0)} + j_+^{(1)} + j_+^{(2)} + j_+^{(3)}  ,  j_-^{(0)} + j_-^{(1)} + j_-^{(2)} + j_-^{(3)} ] \, ,
\end{align}
which gives rise to the equation
\begin{align}
    0 &= \partial_+ \left( j_-^{(0)} + j_-^{(1)} + j_-^{(2)} + j_-^{(3)} \right) - \partial_- \left( j_+^{(0)} + j_+^{(1)} + j_+^{(2)} + j_+^{(3)} \right) \nonumber \\
    &\quad + [ j_+^{(0)} , j_-^{(0)}] + [ j_+^{(0)} , j_-^{(1)}] + [ j_+^{(0)} , j_-^{(2)}] + [ j_+^{(0)} , j_-^{(3)}] \nonumber \\
    &\quad + [ j_+^{(1)} , j_-^{(0)}] + [ j_+^{(1)} , j_-^{(1)}] + [ j_+^{(1)} , j_-^{(2)}] + [ j_+^{(1)} , j_-^{(3)}] \nonumber \\
    &\quad + [ j_+^{(2)} , j_-^{(0)}] + [ j_+^{(2)} , j_-^{(1)}] + [ j_+^{(2)} , j_-^{(2)}] +  [ j_+^{(2)} , j_-^{(3)}] \nonumber \\
    &\quad + [ j_+^{(3)} , j_-^{(0)}] + [ j_+^{(3)} , j_-^{(1)}] + [ j_+^{(3)} , j_-^{(2)}] + [ j_+^{(3)} , j_-^{(3)}] \, .
\end{align}
Recalling that $[ \mathfrak{g}_n, \mathfrak{g}_m ] \subset \mathfrak{g}_{( n + m ) \text{ mod } 4}$, we can now separate terms based on the subspace they belong to. For instance, the term $ [ j_+^{(1)}, j_-^{(2)}]$ contributes to the projection of the Maurer-Cartan identity onto the subspace $\mathfrak{g}_3$, and so on. Thus we find four independent equations:
\begin{align}\label{sSSSM_four_maurer_cartan}
    &\mathfrak{g}_0 \; : \; \partial_+ j_-^{(0)} - \partial_- j_+^{(0)} + [ j_+^{(0)} , j_-^{(0)}] + [ j_+^{(1)} , j_-^{(3)}] + [ j_+^{(2)} , j_-^{(2)}] + [ j_+^{(3)} , j_-^{(1)}] = 0 \, , \nonumber \\
    &\mathfrak{g}_1 \; : \; \partial_+ j_-^{(1)} - \partial_- j_+^{(1)} + [ j_+^{(0)} , j_-^{(1)} ] + [ j_+^{(1)} , j_-^{(0)} ] + [ j_+^{(2)} , j_-^{(3)} ] + [ j_+^{(3)} , j_-^{(2)} ] = 0 \, , \nonumber \\
    &\mathfrak{g}_2 \; : \; \partial_+ j_-^{(2)} - \partial_- j_+^{(2)} + [ j_+^{(0)} , j_-^{(2)} ] + [ j_+^{(2)} , j_-^{(0)} ] + [ j_+^{(1)} , j_-^{(1)} ] + [ j_+^{(3)} , j_-^{(3)} ] = 0  \, , \nonumber \\
    &\mathfrak{g}_3 \; : \; \partial_+ j_-^{(3)} - \partial_- j_+^{(3)} + [ j_+^{(0)} , j_-^{(3)} ] + [ j_+^{(3)} , j_-^{(0)} ] + [ j_+^{(1)} , j_-^{(2)} ] + [ j_+^{(2)} , j_-^{(1)} ] = 0 \, .
\end{align}
Just as before, we define the field strength and covariant derivative associated with $j^{(0)}$, 
\begin{align}
    F_{\alpha \beta}^{(0)} = \partial_\alpha j_\beta^{(0)} - \partial_\beta j_\alpha^{(0)} + [ j_\alpha^{(0)} , j_\beta^{(0)} ] \, , \qquad D_\alpha = \partial_\alpha + \left[ j_\alpha^{(0)} \; , \; \cdot \, \right] \, ,
\end{align}
and in terms of these objects one can rewrite (\ref{sSSSM_four_maurer_cartan}) as
\begin{align}\label{sSSSM_four_maurer_cartan_nicer}
    &\mathfrak{g}_0 \; : \; F_{+-}^{(0)} + [ j_+^{(1)} , j_-^{(3)}] + [ j_+^{(2)} , j_-^{(2)}] + [ j_+^{(3)} , j_-^{(1)}] = 0 \, , \nonumber \\
    &\mathfrak{g}_1 \; : \; D_+ j_-^{(1)} - D_- j_+^{(1)} + [ j_+^{(2)} , j_-^{(3)} ] + [ j_+^{(3)} , j_-^{(2)} ] = 0 \, , \nonumber \\
    &\mathfrak{g}_2 \; : \; D_+ j_-^{(2)} - D_- j_+^{(2)} + [ j_+^{(1)} , j_-^{(1)} ] + [ j_+^{(3)} , j_-^{(3)} ] = 0  \, , \nonumber \\
    &\mathfrak{g}_3 \; : \; D_+ j_-^{(3)} - D_- j_+^{(3)} + [ j_+^{(1)} , j_-^{(2)} ] + [ j_+^{(2)} , j_-^{(1)} ] = 0 \, .
\end{align}
Note that, when the fermionic components $j^{(1)}$ and $j^{(3)}$ are set to zero, this reduces to the set of two equations (\ref{SSSM_two_maurer_cartan}) that we had in the symmetric space case.

\subsubsection*{\ul{\it sSSSM with WZ Term}}

As in the discussion of the PCM with WZ term in Section \ref{sec:AFSM_WZ}, let us now restore the factors of $\hay$ and $\kay$. The semi-symmetric space sigma model with Wess-Zumino term, or sSSSM-WZ, is described by the action
\begin{align}\label{semi_plus_wz}
    S &= \frac{\hay}{2} \int_{\Sigma} \mathrm{str} \left( \eta^{\alpha \beta} j_\alpha^{(2)} j_\beta^{(2)} + \ell \epsilon^{\alpha \beta} j_\alpha^{(1)} j_\beta^{(3)} \right)  + \frac{\kay}{3} \int_{\mathcal{M}_3} \epsilon^{ijk} \mathrm{str} \left( j^{(2)}_i [ j^{(2)}_j , j^{(2)}_k ] + 3 j^{(1)}_i [ j^{(3)}_j , j^{(2)}_k ] \right) \, ,
\end{align}
where $\mathcal{M}_3$ is a $3$-manifold whose boundary is the worldsheet, $\partial \mathcal{M}_3 = \Sigma$, as before. Our conventions are $\eta^{+-} = \eta^{-+} = - \frac{1}{2}$, $\eta_{+-} = \eta_{-+} = - 2$, and $\epsilon^{+-} = - \epsilon^{-+} = \frac{1}{2}$. 

Note that there is a new parameter $\ell$ which multiplies the fermionic term, which was not present in the sSSSM without Wess-Zumino term. The model turns out to be integrable only for the particular value
\begin{align}\label{ell_constraint}
    \ell^2 = 1 - \frac{\kay^2}{\hay^2} \, ,
\end{align}
which reduces to $1$ when $\kay = 0$. See the comments around equation (4.6) of \cite{Cagnazzo:2012se} for a discussion of this fact; from the perspective of the Green-Schwarz superstring, one can also interpret (\ref{ell_constraint}) as the condition for kappa symmetry to be preserved. 

The sSSSM-WZ is also classically integrable, as the Euler-Lagrange equations which arise from the variation of (\ref{semi_plus_wz}) are equivalent to the flatness of the Lax connection
\begin{align}\label{undeformed_sSSSM_WZ_lax}
    \mathfrak{L}_{\pm} &= j_{\pm}^{(0)} + \ell \frac{z^2 + 1}{z^2 - 1} j_{\pm}^{(2)} \pm \left( \frac{\kay}{\hay} - \frac{2 \ell z}{z^2 - 1} \right) j_{\pm}^{(2)} + \left( z + \frac{\ell}{1 - \frac{\kay}{\hay}} \right) \sqrt{ \frac{\ell \left( 1 - \frac{\kay}{\hay} \right)}{z^2 - 1} } j_{\pm}^{(1)} \nonumber \\
    &\qquad + \left( z - \frac{\ell}{1 + \frac{\kay}{\hay}} \right) \sqrt{ \frac{\ell \left( 1 + \frac{\kay}{\hay} \right) }{z^2 - 1} } j_{\pm}^{(3)} \, .
\end{align}

\subsection{sSSSM-WZ with Auxiliary Fields}

We are now in a position to define the semi-symmetric space sigma model with Wess-Zumino term deformed by auxiliary field interactions, which we will refer to by the (admittedly awkward) acronym AF-sSSSM-WZ. Following the strategy of the SSSM, we again introduce auxiliary fields $v_\alpha \in \mathfrak{g}$ which enjoy a decomposition into $v_\alpha^{(n)}$ similar to (\ref{four_subspace_decomp}). Only the projection $v_\alpha^{(2)}$ onto $\mathfrak{g}_2$ will be used to build our auxiliary field couplings, which will be expressed in terms of the combinations
\begin{align}
    \nu_n = \str \Big( \underbrace{ v^{(2)}_+ \ldots v^{(2)}_+}_{n \text{ times}} \Big) \str \Big( \underbrace{ v^{(2)}_- \ldots v^{(2)}_-}_{n \text{ times}} \Big) \, ,
\end{align}
precisely as in the preceding cases. We define the deformed model via the action
\begin{align}\label{semi_plus_wz_deformed}
    S &= \hay \int_{\Sigma}\mathrm{str} \left( \left( \frac{1}{2}  j_+^{(2)} j_-^{(2)} + v_+^{(2)} v_-^{(2)}  + j_+^{(2)} v_-^{(2)} + j_-^{(2)} v_+^{(2)} \right) + \ell x_0 + E ( \nu_2 , \ldots , \nu_N ) \right) \nonumber \\
    &\qquad + \frac{\kay}{3} \int_{\mathcal{M}_3} \epsilon^{ijk} \mathrm{str} \left( j^{(2)}_i [ j^{(2)}_j , j^{(2)}_k ] + 3 j^{(1)}_i [ j^{(3)}_j , j^{(2)}_k ] \right) \, ,
\end{align}
with $x_0$ as in (\ref{x0_defn}). 
Note that, as was already mentioned in the previous sections, there are case-by-case constraints on the set of independent variables $\nu_n$. Here we are interested in model-independent structures, and we refer the reader to, e.g., \cite{Evans:1999mj,Evans:2000qx,Evans:2000hx,Evans:2001sz,Evans:2004mu,Miller:2006bu,Komatsu:2019hgc}, for various restrictions that can arise on the independent invariant tensors and higher-spin currents and, therefore, on the $\nu_n$ variables.

A new subtlety of the semi-symmetric case is the following. In our analysis of the SSSM, we used the fact that $\tr \left( X^{(n)} X^{(m)} \right) = 0$ unless $n = m$, where $X^{(n)} \in \mathfrak{g}_n$ and $X^{(m)} \in \mathfrak{g}_m$. However, for the sSSSM-WZ, note that the Wess-Zumino term in (\ref{semi_plus_wz_deformed}) involves supertraces of quantities like $j^{(2)}_i [ j^{(2)}_j , j^{(2)}_k ]$, where one has $j_i^{(2)} \in \mathfrak{g}_2$ and $[ j^{(2)}_j , j^{(2)}_k ] \in \mathfrak{g}_0$. This supertrace need not vanish, despite the fact that its argument is a product of terms that belong to two different subspaces $\mathfrak{g}_n$. Thus when we vary the action to compute the equations of motion, we are no longer permitted to project these equations onto the four independent subspaces $\mathfrak{g}_n$, because the Wess-Zumino term will mix these projections; one says that the action is now incompatible with the $\mathbb{Z}_4$ grading. This construction was first considered in \cite{Cagnazzo:2012se}; see also \cite{Borsato:2022tmu} for further comments. However, we note that the Maurer-Cartan identity is independent of the Wess-Zumino term, so the projections (\ref{sSSSM_four_maurer_cartan_nicer}) of this identity onto the four subspaces $\mathfrak{g}_n$ is still valid.
Furthermore, we can also still use the preserved $\mathbb{Z}_2$ grading in the supertrace, so the supertrace of a product of a boson and a fermion must still vanish.

The equations of motion for the AF-sSSSM-WZ are derived in Appendix \ref{app:sSSSM_eom}. The auxiliary field equation of motion takes the same form as in the AF-SSSM, equation (\ref{SSSM_auxiliary_eom}). Here we will record only the simplified version of the $g$-field equation of motion, after using the Maurer-Cartan identity (\ref{sSSSM_four_maurer_cartan_nicer}). Under this assumption, the $g$-field equation of motion implies the three separate conditions
\begin{align}\label{sSSSM_eom_with_MC}
    0 &\deq  [ \mathfrak{J}^{(2)}_+ , j^{(1)}_- ] \!+\! [ \mathfrak{J}^{(2)}_- , j^{(1)}_+ ] \!-\!  \ell \left( [ j^{(2)}_+ , j^{(1)}_- ] \!-\! [ j^{(2)}_- , j^{(1)}_+ ] \right) \!-\! \frac{\kay}{\hay} \left(  [ j^{(3)}_+  , j^{(2)}_- ] \!-\! [ j^{(3)}_- , j^{(2)}_+  ] \right) \, , \nonumber \\
    0 &\deq D_+ \mathfrak{J}^{(2)}_- \!+\! D_- \mathfrak{J}^{(2)}_+  \!-\! \ell \left( [ j_+^{(3)} , j_-^{(3)} ] \!-\! [ j_+^{(1)} , j_-^{(1)} ] \right) \!+\! \frac{\kay}{\hay} \left( 2 [ j_+^{(2)} , j_-^{(2)} ]  \!+\! [ j_+^{(1)} , j_-^{(3)} ] \!-\! [ j_-^{(1)} , j_+^{(3)} ] \right) \, , \nonumber \\
    0 &\deq [ \mathfrak{J}^{(2)}_+ , j^{(3)}_- ] \!+\! [ \mathfrak{J}^{(2)}_- , j^{(3)}_+ ] \!+\! \ell \left( [ j^{(2)}_+ , j^{(3)}_- ] \!-\! [ j^{(2)}_- , j^{(3)}_+ ] \right) \!-\! \frac{\kay}{\hay} \left( [ j^{(1)}_+ , j^{(2)}_- ] \!-\! [ j^{(1)}_- , j^{(2)}_+  ] \right) \, .
\end{align}
There is a fourth $g$-field equation of motion which is automatically satisfied when the auxiliary field Euler-Lagrange equation holds, so we do not report it. However, since -- strictly speaking -- the full set of the four $g$-field equations are only satisfied when \emph{both} this fourth equation (which is implied by the auxiliary field equation of motion) and the three conditions (\ref{sSSSM_eom_with_MC}) are all obeyed, we have written the conditions (\ref{sSSSM_eom_with_MC}) with the dot-equality symbol $\deq$. This symbol is to be interpreted as the statement that, when the $v_\pm$ equation of motion is satisfied, the full set of $g$-field equations of motion are equivalent to the three conditions (\ref{sSSSM_eom_with_MC}). Similar behavior occurred in the AF-SSSM equation of motion (\ref{SSSM_eom_simplified}), to which (\ref{sSSSM_eom_with_MC}) reduces upon setting $\kay = 0$ and dropping all fermionic terms.

\subsection{Classical Integrability}

We will now verify that the equations of motion for the family of models (\ref{semi_plus_wz_deformed}) are equivalent to the flatness of the Lax connection 
\begin{align}\label{sSSSM_WZ_lax}
    \mathfrak{L}_{\pm} &= j_{\pm}^{(0)} + \ell \frac{z^2 + 1}{z^2 - 1} j_{\pm}^{(2)} \pm \left( \frac{\kay}{\hay} - \frac{2 \ell z}{z^2 - 1 } \right) \mathfrak{J}_{\pm}^{(2)} + \left( z + \frac{\ell}{1 - \frac{\kay}{\hay} } \right) \sqrt{ \frac{ \ell \left( 1 - \frac{\kay}{\hay} \right) }{z^2 - 1 } } j_{\pm}^{(1)} \nonumber \\
    &\quad + \left( z - \frac{\ell}{1 + \frac{\kay}{\hay} } \right) \sqrt{ \frac{ \ell \left( 1 + \frac{\kay}{\hay} \right) }{ z^2 - 1 } } j_{\pm}^{(3)} \, .
\end{align}
As a first check, when $E = 0$ and the auxiliary field has been integrated out using its equation of motion, recall that $\mathfrak{J}_{\pm}^{(2)} \deq j_{\pm}^{(2)}$, and in this limit the candidate Lax connection (\ref{sSSSM_WZ_lax}) manifestly reduces to the Lax (\ref{undeformed_sSSSM_WZ_lax}) of the undeformed sSSSM with WZ term. Next we would like to show that the flatness of this Lax connection is equivalent to the AF-sSSSM equations of motion for any choice of interaction function $E ( \nu_2 , \ldots, \nu_N )$.

The logic of this proof is similar to that of the cases we have discussed in previous sections. The details of the computation of the curvature $d_{\mathfrak{L}} \mathfrak{L} = \partial_+ \mathfrak{L}_- - \partial_- \mathfrak{L}_+ + [ \mathfrak{L}_+ , \mathfrak{L}_- ]$ are presented in Appendix \ref{app:sSSSM_curvature}. Due to the breaking of the $\mathbb{Z}_4$ grading that is implied by the non-zero Wess-Zumino term, we cannot project the curvature of the Lax connection onto each $\mathfrak{g}_n$ separately, but we may still extract its components along the bosonic and fermionic subspaces $\mathfrak{g}_B = \mathfrak{g}_0 \oplus \mathfrak{g}_2$ and $\mathfrak{g}_F = \mathfrak{g}_1 \oplus \mathfrak{g}_3$. The bosonic contribution is given in equation (\ref{bosonic_sSSSM_lax_final}), which we reproduce here:
\begin{align}\label{bosonic_sSSSM_lax_final_body}
    \left( d_{\mathfrak{L}} \mathfrak{L} \right) \big\vert_{\mathfrak{g}_B} \deq - \left( \frac{\kay}{\hay} - \frac{2 \ell z}{z^2 - 1} \right) &\Bigg( D_- \mathfrak{J}_+^{(2)} + D_+ \mathfrak{J}_-^{(2)} + \frac{\kay}{\hay} \left( 2 [ j_+^{(2)} , j_-^{(2)} ] +  [ j_+^{(1)} , j_-^{(3)} ] + [ j_+^{(3)} , j_-^{(1)} ] \right) \nonumber \\
    &\qquad \qquad - \ell \left( [ j_+^{(3)} , j_-^{(3)} ] - [ j_+^{(1)} , j_-^{(1)} ] \right) \Bigg) \, .
\end{align}
The quantity in parentheses in (\ref{bosonic_sSSSM_lax_final_body}) is exactly the second line of the equation of motion (\ref{sSSSM_eom_with_MC}). Therefore, we find that the bosonic part of the Lax is flat if and only if the second equation of motion is satisfied.

Next let us consider the fermionic part of the flatness condition, which is obtained in equation (\ref{fermion_final}). To ease notation, let us define coefficients
\begin{gather}
    c_1 \!=\! \left( z \!+\! \frac{\ell}{1 \!-\! \frac{\kay}{\hay} } \right) \sqrt{ \frac{ \ell \left( 1 \!-\! \frac{\kay}{\hay} \right) }{z^2 \!-\! 1 } } \, , \quad 
    c_2 \!=\! \left( z \!-\! \frac{\ell}{1 \!+\! \frac{\kay}{\hay} } \right) \sqrt{ \frac{ \ell \left( 1 \!+\! \frac{\kay}{\hay} \right) }{ z^2 \!-\! 1 } } \, , \nonumber \\ 
    c_3 \!=\! \ell \frac{z^2 \!+\! 1}{z^2 \!-\! 1} \left( z \!+\! \frac{\ell}{1 \!-\! \frac{\kay}{\hay} } \right) \sqrt{ \frac{ \ell \left( 1 \!-\! \frac{\kay}{\hay} \right) }{z^2 \!-\! 1 } } \, , \quad  
    c_4 \!=\! \ell \frac{z^2 \!+\! 1}{z^2 \!-\! 1} \left( z \!-\! \frac{\ell}{1 \!+\! \frac{\kay}{\hay} } \right) \sqrt{ \frac{ \ell \left( 1 \!+\! \frac{\kay}{\hay} \right) }{ z^2 \!-\! 1 } } \, ,  \\ 
    \hspace{-15pt} c_5 \!=\! \left( \frac{\kay}{\hay} \!-\! \frac{2 \ell z}{z^2 \!-\! 1 } \right) \left( z \!+\! \frac{\ell}{1 \!-\! \frac{\kay}{\hay} } \right) \sqrt{ \frac{ \ell \left( 1 \!-\! \frac{\kay}{\hay} \right) }{z^2 \!-\! 1 } }  \, , \quad 
    c_6 \!=\! \left( \frac{\kay}{\hay} \!-\! \frac{2 \ell z}{z^2 \!-\! 1 } \right) \left( z \!-\! \frac{\ell}{1 \!+\! \frac{\kay}{\hay} } \right) \sqrt{ \frac{ \ell \left( 1 \!+\! \frac{\kay}{\hay} \right) }{ z^2 \!-\! 1 } } \, , \notag
\end{gather}
which we remember are still functions of $z$, so that (\ref{fermion_final}) can be written as
\begin{align}\label{simplified_fermionic_flatness}
    \left( d_\mathfrak{L} \mathfrak{L} \right) \big\vert_{\mathfrak{g}_F} &= \left( c_4 - c_1 \right) \left( [ j_+^{(2)} , j_-^{(3)} ] + [ j_+^{(3)} , j_-^{(2)} ] \right) + \left( c_3 - c_2 \right) \left( [ j_+^{(1)} , j_-^{(2)} ] + [ j_+^{(2)} , j_-^{(1)} ] \right) \nonumber \\
    &\quad + c_5 \left( [ \mathfrak{J}_+^{(2)} , j_-^{(1)} ] + [ \mathfrak{J}_-^{(2)} , j_+^{(1)} ] \right) + c_6 \left( [ \mathfrak{J}_+^{(2)} , j_-^{(3)} ] + [ \mathfrak{J}_-^{(2)} , j_+^{(3)} ] \right) \, .
\end{align}
Note that the auxiliary field equations of motion have not been used in computing the fermionic truncation of the curvature, so we have written (\ref{simplified_fermionic_flatness}) with the true equality symbol $=$ rather than dot-equality $\deq$. However, this does not affect the logic of the proof of weak integrability; two quantities which are truly equal are also equal when the auxiliary field equations of motion are satisfied.

To show that the remaining (fermionic) equations of motion are equivalent to the flatness of the fermionic projection (\ref{simplified_fermionic_flatness}), we must prove both implications. First assume that the equations of motion (\ref{sSSSM_eom_with_MC}) are satisfied. When this is true, the flatness condition becomes
\begin{align}
    \left( d_\mathfrak{L} \mathfrak{L} \right) \big\vert_{\mathfrak{g}_F} &= \left( c_4 - c_1 - \ell c_6 + \frac{\kay}{\hay} c_5 \right) \left( [ j_+^{(2)} , j_-^{(3)} ] + [ j_+^{(3)} , j_-^{(2)} ] \right) \nonumber \\
    &\quad + \left( c_3 - c_2 + \ell c_5 + \frac{\kay}{\hay} c_6 \right) \left( [ j_+^{(1)} , j_-^{(2)} ] + [ j_+^{(2)} , j_-^{(1)} ] \right) \, .
\end{align}
However, by a direct computation, one finds that
\begin{equation}
\begin{aligned}
     0 &= c_4 - c_1 - \ell c_6 + \frac{\kay}{\hay} c_5 \, , 
     \\
     0 &= c_3 - c_2 + \ell c_5 + \frac{\kay}{\hay} c_6 \, ,
\end{aligned}
\end{equation}
assuming the relation (\ref{ell_constraint}). This proves one direction.

On the other hand, suppose that (\ref{simplified_fermionic_flatness}) vanishes. This gives one equation for each value of $z$; we choose to evaluate this condition at the two points
\begin{align}
    z = \pm \frac{\ell}{1 \pm \frac{\kay}{\hay}} \, ,
\end{align}
giving two independent equations, which admit the unique simultaneous solution
\begin{equation}
\begin{aligned}
    0 &\!=\! [ \mathfrak{J}^{(2)}_+ , j^{(1)}_- ] \!+\! [ \mathfrak{J}^{(2)}_- , j^{(1)}_+ ] \!-\!  \ell \left( [ j^{(2)}_+ , j^{(1)}_- ] \!-\! [ j^{(2)}_- , j^{(1)}_+ ] \right) \!-\! \frac{\kay}{\hay} \left(  [ j^{(3)}_+  , j^{(2)}_- ] \!-\! [ j^{(3)}_- , j^{(2)}_+  ] \right) \, ,
    \\
    0 &\!=\! [ \mathfrak{J}^{(2)}_+ , j^{(3)}_- ] \!+\! [ \mathfrak{J}^{(2)}_- , j^{(3)}_+ ] \!+\! \ell \left( [ j^{(2)}_+ , j^{(3)}_- ] \!-\! [ j^{(2)}_- , j^{(3)}_+ ] \right) \!-\! \frac{\kay}{\hay} \left( [ j^{(1)}_+ , j^{(2)}_- ] \!-\! [ j^{(1)}_- , j^{(2)}_+  ] \right) \, ,
\end{aligned}
\end{equation}
which are the first and third lines of (\ref{sSSSM_eom_with_MC}). This establishes the reverse direction, so we conclude that every member of this family of AF-sSSSM-WZ models is weakly classically integrable, with a Lax connection given by (\ref{sSSSM_WZ_lax}).

\section{Conclusion}\label{sec:conclusion}

In this work, we have constructed infinitely many integrable deformations of any seed sigma model whose target space is a semi-symmetric coset, and whose action may include a Wess-Zumino term. These integrable deformations are in one-to-one correspondence with functions $E ( \nu_2, \ldots , \nu_N )$ of several variables. This construction generalizes several of the other auxiliary field deformations which have appeared in recent works. For instance, truncating these deformed models by setting fermions to zero and focusing on a coset $( G \times G ) / G$ recovers the higher-spin auxiliary field deformation of the PCM with Wess-Zumino term of \cite{Bielli:2024ach}. Another interesting special case is to set both the fermions and the WZ term to zero, but to consider general symmetric cosets, which we studied in Section \ref{sec:SSSM}.

There remain several interesting directions for future research. Perhaps the most exciting is to investigate particular cases of semi-symmetric space sigma models which describe string propagation on spacetimes such as $\mathrm{AdS}_5 \times S^5$ or $\mathrm{AdS}_3 \times S^3 \times T^4$, as such models have provided much of the motivation for studies of integrable deformations in recent years.

Below we outline two other lines of inquiry, which we hope to return to in future work.

\subsubsection*{\ul{\it Closed-form solutions for higher-spin flow equations}}

In the case of an interaction function $E ( \nu_2 )$ which depends on only the single variable $\nu_2 = \tr ( v_+ v_+ ) \tr ( v_- v_- )$, the auxiliary field deformations of interest in this article include all flows passing through the seed theory which are driven by functions of the energy-momentum tensor. For instance, this class of deformations includes the $\TT$ \cite{Zamolodchikov:2004ce,Cavaglia:2016oda} and root-$\TT$ \cite{Ferko:2022cix,Conti:2022egv,Babaei-Aghbolagh:2022uij,Babaei-Aghbolagh:2022leo} flows. There has been considerable recent interest in solving classical flow equations driven by such functions of the energy-momentum tensor in diverse spacetime dimensions, both in $2d$ field theories where such flows can be defined at the quantum level \cite{Cavaglia:2016oda,Bonelli:2018kik,Brennan:2019azg}, but also for $1d$ mechanical models \cite{Gross:2019ach,Gross:2019uxi,Chakraborty:2020xwo,Ferko:2023ozb,Giordano:2023wgx,Ferko:2023iha}, three-dimensional models \cite{Ferko:2023sps}, $4d$ theories of electrodynamics \cite{Conti:2018jho,Babaei-Aghbolagh:2020kjg,Babaei-Aghbolagh:2022uij,Ferko:2023ruw,Ferko:2023wyi,Babaei-Aghbolagh:2024uqp}, and for $6d$ chiral tensor theories \cite{Ferko:2024zth}. Other examples include classical flows in cases with supersymmetry \cite{Baggio:2018rpv,Chang:2018dge,Chang:2019kiu,Coleman:2019dvf,Jiang:2019hux,Jiang:2019trm,Ferko:2019oyv,Lee:2021iut,Ferko:2021loo,Lee:2023uxj}, or sequential flows by multiple $\TT$ deformations \cite{Ferko:2022dpg}. In many cases, such classical\footnote{In some cases, \emph{quantum} aspects of root-$\TT$ deformed theories have been investigated perturbatively, such as for $2d$ bosons or $4d$ gauge theory \cite{Ebert:2024zwv,LukeMartin:2024gsb}; the latter deformation is related to the $4d$ ModMax theory \cite{Bandos:2020jsw}. See \cite{Hadasz:2024pew} for a different approach to defining the root-$\TT$ deformation at the quantum level.} flows can be presented in a geometrical language \cite{Conti:2018tca,Tolley:2019nmm,Conti:2022egv,Morone:2024ffm,Babaei-Aghbolagh:2024hti,Tsolakidis:2024wut,Ferko:2024yhc}; see \cite{Brizio:2024arr} for a recent review of this perspective.

On the other hand, classical flow equations driven by higher-spin conserved currents are comparatively poorly understood. Although there has been progress in understanding so-called $T \overbar{T}_s$ flows \cite{Conti:2019dxg}, built from both the stress tensor and higher-spin currents, it will be important to see whether closed-form expressions can be obtained for the Lagrangian deformed by a flow driven by a combination like $T_s \overbar{T}_s$, where $T_s$ and $\overbar{T}_s$ are spin-$s$ conserved currents in a $2d$ IQFT. One might expect that the higher-spin auxiliary field formalism developed here and in related previous works might be a useful tool in addressing this question. For instance, if one could obtain explicit formulas for the spin-$s$ conserved currents in an auxiliary field sigma model, and then study the differential equation
\begin{align}
    \frac{\partial E}{\partial \lambda} \deq T_s \overbar{T}_s \, ,
\end{align}
possibly using simplifications afforded by the auxiliary field equation of motion, it may be possible to find a closed-form expression for the resulting interaction function $E$, at least in certain cases like $s = 3$. Integrating out the auxiliary fields would then yield an expression for the deformed Lagrangian, which would provide new examples of classical solutions to flows driven by higher-spin operators of Smirnov-Zamolodchikov type \cite{Smirnov:2016lqw}.

\subsubsection*{\ul{\it Connections between $2d$ and $4d$; T-duality and S-duality}}

As we have reviewed, the structure of the auxiliary field deformations of $2d$ sigma models which we studied in this work is inspired by the Ivanov-Zupnik formalism for theories of duality-invariant electrodynamics in four dimensions \cite{Ivanov:2002ab,Ivanov:2003uj}. This electric-magnetic duality invariance, which is incorporated in this formulation via the introduction of similar auxiliary fields, is a special case of S-duality for $4d$ gauge theories where the gauge group is $U(1)$.

There are many interesting connections between S-duality for $4d$ gauge theories and T-duality for $2d$ sigma models. For instance, both can be realized as canonical transformations which have very similar forms for their generating functions \cite{Alvarez:1994wj,Lozano:1995aq,Lozano:1995jx,Lozano:1996sc}. One way to understand this observation is to note that a particular dimensional reduction of $4d$ gauge theories to two spacetime dimensions maps S-duality onto T-duality \cite{Bershadsky:1995vm,Harvey:1995tg} (see also \cite{McLoughlin:2024ldp} for recent related work in the context of celestial amplitudes).

Although the interplay between non-Abelian T-duality and auxiliary field deformations of $2d$ sigma models has been recently considered \cite{Bielli:2024khq,Bielli:2024ach}, it would be very interesting to see whether a $4d$ analogue of this analysis can be developed which involves $S$-duality. As a first step, such an enterprise would likely require the development of a version of the Ivanov-Zupnik formalism for non-Abelian gauge theories. If this can be achieved, it may well be that a dimensional reduction of this formalism would be closely related to auxiliary field deformations of $2d$ integrable sigma models that we have generalized in this work.

\section*{Acknowledgements}

We thank Mattia Cesaro, Ben Hoare, Zejun Huang, Fiona Seibold, Alessandro Sfondrini, and Konstantin Zarembo for helpful discussions. 
We are also grateful to Mattia Cesaro, Axel Kleinschmidt, and David Osten for contacting us on September 2nd 2024 to announce the submission of their paper \cite{Cesaro:2024ipq}, and then coordinating with us to jointly submit the two preprints to the arXiv.
C.\,F. and G.\,T.-M. are grateful to the participants of the meeting ``Integrability in low-supersymmetry theories,'' held in Trani in 2024 
and funded by the COST Action CA22113 by INFN and by Salento University, for stimulating discussions on topics related to the subject of this work.
D.\,B. is supported by Thailand NSRF via PMU-B, grant number B13F670063.
C.\,F. is supported by the National Science Foundation under Cooperative Agreement PHY-2019786 (the NSF AI Institute for Artificial Intelligence and Fundamental Interactions).
L.\,S. is supported by a postgraduate scholarship at the University of Queensland.
G.\,T.-M. has been supported by the Australian Research Council (ARC) Future Fellowship FT180100353, ARC Discovery
Project DP240101409, and the Capacity Building Package of the University of Queensland.

\appendix


\section{Details of SSSM Calculations}\label{app:SSSM}

In this Appendix, we collect some of the more involved calculations which are used in the discussion of Section \ref{sec:SSSM}, in order to streamline the presentation in the body of this article.

\subsection{Equations of Motion}\label{app:SSSM_eom}

We begin by deriving the equations of motion associated with the deformed symmetric space sigma model (\ref{SSSM_deformed}). As usual, there is one equation of motion associated with $v_{\pm}$ and one equation of motion for $g$.

Let us begin with the auxiliary field equation of motion. When we decompose the field $v_\alpha$ according to (\ref{SSSM_aux_decomp}), we see that the $v^{(0)}_\alpha$ (the projection of $v_\alpha$ onto $\mathfrak{g}_0$) does not appear anywhere in the Lagrangian. Therefore one has
\begin{align}
    \frac{\partial \mathcal{L}}{\partial v_{\pm}} &= \frac{\partial \mathcal{L}}{\partial v_{\pm}^{(2)}} \, .
\end{align}
Let us stress that this behavior is different from that of the physical field $g$ and associated Maurer-Cartan form $j_\alpha$, since in the latter case the equation of motion is obtained by varying the group-valued field $g$, whereas the auxiliary field $v_\alpha$ is Lie algebra valued.

As a consequence, the Euler-Lagrange equation for the auxiliary field takes the same form as it did in the deformed PCM-like models,
\begin{align}\label{SSSM_auxiliary_eom_app}
    0 &\deq j_{\pm}^{(2)} + v_{\pm}^{(2)} \\
    &\quad + \sum_{n = 2}^{N} n \frac{\partial E}{\partial \nu_n} \tr \left( \Big( v_{\pm}^{(2)} \Big)^n \right) \Big( v_{\mp}^{(2)} \Big)^{A_1} \left( v_{\mp}^{(2)} \right)^{A_2} \ldots \left( v_{\mp}^{(2)} \right)^{A_{n-1}} T^{A_n} \tr ( T_{(A_1} T_{A_2} \ldots T_{A_{n} )} ) \, . \notag
\end{align}
The only difference between the equation of motion (\ref{SSSM_auxiliary_eom_app}) in the AF-SSSM and the corresponding Euler-Lagrange equation (\ref{v_eom}) for the AFSM is that $j_\pm$ and $v_\pm$ have been replaced by their projections onto $\mathfrak{g}_2$.

In contrast, the equation of motion for $j_\pm$ will be modified in a more involved way compared to the AFSM. However, the derivation of this equation follows essentially the same steps as in the usual symmetric space sigma model. Consider right-multiplication of the group-valued field $g$ by any $e^{\epsilon} \in G$:
\begin{align}
    g \longrightarrow g e^{\epsilon} \approx g (1 + \epsilon)  \, , 
\end{align}
where $\epsilon \in \mathfrak{g}$, so that $\delta g = g \epsilon$ and $\delta g^{-1} = - \epsilon g^{-1}$. Under such a variation, the left-invariant Maurer-Cartan form varies as
\begin{align}
    \delta j_\pm &= \partial_\pm \epsilon + [ j_\pm , \epsilon ] =  \partial_\pm \epsilon + [ j^{(0)}_{\pm}  , \epsilon ] + [  j^{(2)}_{\pm}  , \epsilon ] \, .
\end{align}
In terms of the covariant derivative defined in (\ref{SSSM_covariant_deriv}), this is
\begin{align}\label{SSSM_delta_j}
    \delta j_{\pm} = D_\pm \epsilon + [ j^{(2)}_{\pm} , \epsilon ] \, .
\end{align}
Now consider the variation of each $j_\pm$-dependent term in the Lagrangian (\ref{SSSM_deformed}). First,
\begin{align}
    \delta \left( \tr ( j_+^{(2)} j_-^{(2)} ) \right) &= \tr \left( j_+^{(2)} \delta j_-^{(2)}  +  j_-^{(2)} \delta j_+^{(2)} \right) \nonumber \\
    &= \tr \left( j_+^{(2)} \left( \delta j_- - \delta j_-^{(0)} \right)  +  j_-^{(2)}  \left( \delta j_+ - \delta j_+^{(0)} \right)  \right) \nonumber \\
    &= \tr \left( j_+^{(2)} \delta j_-  +  j_-^{(2)} \delta j_+  \right) \, ,
\end{align}
where we have used that
\begin{align}\label{vanish_orthogonality}
    \tr ( j_+^{(2)} \delta j_-^{(0)} ) = \tr ( j_-^{(2)} \delta j_+^{(0)} ) = 0 \, ,
\end{align}
since $j_{\pm}^{(2)} \in \mathfrak{g}_2$ while $\delta j_{\pm}^{(0)} \in \mathfrak{g}_0$, which are orthogonal subspaces, so the trace of the products in (\ref{vanish_orthogonality}) must vanish by orthogonality. Substituting for $\delta j_{\pm}$ using (\ref{SSSM_delta_j}) then gives
\begin{align}\label{SSSM_variation_first_term}
    \delta \left( \tr ( j_+^{(2)} j_-^{(2)} ) \right) &= \tr \left( j_+^{(2)} \left( D_- \epsilon + [ j_-^{(2)} , \epsilon ] \right) +  j_-^{(2)} \left( D_+ \epsilon + [ j_+^{(2)} , \epsilon ] \right) \right) \nonumber \\
    &= \tr \left( j_+^{(2)} D_- \epsilon + j_-^{(2)} D_+ \epsilon \right) \, ,
\end{align}
where in the last step we have used that the commutators cancel by cyclicity, i.e.
\begin{align}
    \tr \left( j_+^{(2)} [ j_-^{(2)} , \epsilon ]  +  j_-^{(2)} [ j_+^{(2)} , \epsilon ] \right) &= \tr \left( \epsilon [ j_+^{(2)} , j_-^{(2)} ] + \epsilon [ j_-^{(2)} , j_+^{(2)} ] \right) = 0 \, .
\end{align}
The steps in varying the coupling term are similar. One obtains
\begin{align}\label{SSSM_coupling_variation}
    \delta \left( \tr ( j_+^{(2)} v_-^{(2)} + j_-^{(2)} v_+^{(2)} ) \right) &= \tr \left( \delta j_+^{(2)} v_-^{(2)} + \delta j_-^{(2)} v_+^{(2)} \right) \nonumber \\
    &= \tr \left( \left( \delta j_+ - \delta j_+^{(0)} \right) v_-^{(2)} + \left( \delta j_- - \delta j_-^{(0)} \right) v_+^{(2)} \right) \nonumber \\
    &= \tr \left( \left( D_+ \epsilon + [ j_+^{(2)} , \epsilon ] \right) v_-^{(2)} + \left( D_- \epsilon + [ j_-^{(2)} , \epsilon ] \right) v_+^{(2)} \right) \, ,
\end{align}
where again in going to the final line we have used that $\tr ( \delta j_{\pm}^{(0)} v_{\mp}^{(2)} ) = 0$ by orthogonality.

We now combine the terms (\ref{SSSM_variation_first_term}) and (\ref{SSSM_coupling_variation}) and restore the spacetime integral to obtain the variation of the action rather than the Lagrangian, which gives
\begin{align}
    \delta S \!=\! \!\int_{\Sigma} \! d^2 \sigma \, \tr \left( \frac{1}{2} j_+^{(2)} D_- \epsilon \!+\! \frac{1}{2} 
 j_-^{(2)} D_+ \epsilon \!+\! \left( D_+ \epsilon \!+\! [ j_+^{(2)} , \epsilon ] \right) v_-^{(2)} \!+\! \left( D_- \epsilon \!+\! [ j_-^{(2)} , \epsilon ] \right) v_+^{(2)} \right) \, .
\end{align}
Next one can integrate all of the covariant derivatives $D_{\pm}$ by parts and use the identity
\begin{align}\label{trace_commutator_identity}
    \tr \left( [ j_+^{(2)} , \epsilon ] v_{-}^{(2)} + [ j_-^{(2)} , \epsilon ] v_+^{(2)} \right) = \tr \left( \epsilon \left( [ v_-^{(2)} , j_+^{(2)} ] + [ v_+^{(2)}, j_-^{(2)} ] \right) \right) \, ,
\end{align}
so that the variation of the action can be written as
\begin{align}
    \delta S \!=\! - \frac{1}{2} \!\int_{\Sigma}\! \!d^2\! \sigma  \tr \!\Biggl(\! \!\epsilon\! \left( \!D_- j_+^{(2)} \!+\! D_+ j_-^{(2)} \!+\! 2 D_- v_+^{(2)} \!+\! 2 D_+ v_-^{(2)} \!-\! 2 [ v_-^{(2)} , j_+^{(2)} ] \!-\! 2 [ v_+^{(2)} , j_-^{(2)} ] \right) \!\Biggr)\! \, . 
\end{align}
We conclude that the equation of motion is
\begin{align}
    D_- j_+^{(2)} + D_+ j_-^{(2)} + 2 D_- v_+^{(2)} + 2 D_+ v_-^{(2)} - 2 [ v_-^{(2)} , j_+^{(2)} ] - 2 [ v_+^{(2)} , j_-^{(2)} ] = 0 \, ,
\end{align}
which can also be written as
\begin{align}\label{SSSM_eom_unsimplified_app}
    D_- \left( j_+^{(2)} + 2 v_+^{(2)} \right) + D_+ \left( j_-^{(2)} + 2 v_-^{(2)} \right) = 2 \left(   [ v_-^{(2)} , j_+^{(2)} ] + [ v_+^{(2)} , j_-^{(2)} ] \right) \, .
\end{align}

\subsection{Curvature of Lax Connection}\label{app:SSSM_lax}

Let us now describe some of the steps which lead to the result (\ref{final_AFSSSM_lax}) for the curvature of the Lax connection of the SSSM deformed by auxiliary fields, which we quote in Section \ref{sec:SSSM}.

First, we have seen that when the auxiliary field equation of motion is satisfied, one has the commutator identity (\ref{sssm_fundamental_commutator}). This relation implies that
\begin{align}\label{comp_one}
    [ \mathfrak{J}_+^{(2)} , j_-^{(2)} ] = - [ j_+^{(2)} + 2 v_+^{(2)} , j_-^{(2)} ] \deq - [ j_+^{(2)} , j_-^{(2)} ] - 2 [ v_-^{(2)} , j_+^{(2)} ] \, ,
\end{align}
and
\begin{align}\label{comp_two}
     [ j_+^{(2)} , \mathfrak{J}_-^{(2)} ] = - [ j_+^{(2)} , j_+^{(2)} + 2 v_-^{(2)} ] \deq - [ j_+^{(2)} , j_-^{(2)} ] - 2 [ v_-^{(2)} , j_+^{(2)} ] \, ,
\end{align}
and comparing (\ref{comp_one}) with (\ref{comp_two}) yields
\begin{align}\label{justify_one}
    [ \mathfrak{J}_+^{(2)} , j_-^{(2)} ] \deq [ j_+^{(2)} , \mathfrak{J}_-^{(2)} ] \, .
\end{align}
Similarly, one has
\begin{align}\label{justify_two}
    [ \mathfrak{J}_+^{(2)} , \mathfrak{J}_-^{(2)} ] &= [ j_+^{(2)} + 2 v_+^{(2)} , j_-^{(2)} + 2 v_-^{(2)} ] \nonumber \\
    &\deq [ j_+^{(2)} , j_-^{(2)} ] + 2 [ v_-^{(2)} , v_+^{(2)} ] + 2 [ v_-^{(2)} , v_+^{(2)} ] + 4 [ v_+^{(2)} , v_-^{(2)} ] \nonumber \\
    &= [ j_+^{(2)} , j_-^{(2)} ] \, .
\end{align}
This justifies the commutator relations (\ref{afsssm_derived_commutators}) quoted in Section \ref{sec:afsssm_integrability} of the article. These formulas are useful in evaluating the commutator $\left[ \mathfrak{L}_+ , \mathfrak{L}_- \right]$. One has
\begin{align}
    \left[ \mathfrak{L}_+ , \mathfrak{L}_- \right] &= \left[ j_+^{(0)} + \frac{ \left( z^2 + 1 \right) j_{+}^{(2)} - 2 z \mathfrak{J}_{+}^{(2)}}{z^2 - 1} , j_-^{(0)} + \frac{ \left( z^2 + 1 \right) j_{-}^{(2)} + 2 z \mathfrak{J}_{-}^{(2)}}{z^2 - 1} \right] \nonumber \\
    &= [ j_+^{(0)} , j_-^{(0)} ] + \frac{1}{z^2 - 1} \left( \left[ j_+^{(0)} , ( z^2 + 1 ) j_-^{(2)} + 2 z \mathfrak{J}_-^{(2)} \right] + \left[ ( z^2 + 1 ) j_+^{(2)} - 2 z \mathfrak{J}_+^{(2)} , j_-^{(0)} \right] \right) \nonumber \\
    &\qquad + \frac{1}{ ( z^2 - 1 )^2} \left( \left[ ( z^2 + 1 ) j_+^{(2)} - 2 z \mathfrak{J}_+^{(2)} , ( z^2 + 1 ) j_-^{(2)} + 2 z \mathfrak{J}_-^{(2)}   \right] \right) \, .
\end{align}
The quantity in parentheses on the final line is
\begin{align}
    &\left[ ( z^2 + 1 ) j_+^{(2)} - 2 z \mathfrak{J}_+^{(2)} , ( z^2 + 1 ) j_-^{(2)} + 2 z \mathfrak{J}_-^{(2)}   \right] \nonumber \\
    &\qquad = ( z^2 + 1 )^2 [ j_+^{(2)} , j_-^{(2)} ] - 2 z ( z^2 + 1 ) \left( [ \mathfrak{J}_+^{(2)} , j_-^{(2)} ] - [ j_+^{(2)} , \mathfrak{J}_-^{(2)} ] \right) - 4 z^2 [ \mathfrak{J}_+^{(2)} , \mathfrak{J}_-^{(2)} ] \nonumber \\
    &\qquad \deq ( z^2 + 1 )^2 [ j_+^{(2)} , j_-^{(2)} ] - 4 z^2 [ j_+^{(2)} , j_-^{(2)} ] \nonumber \\
    &\qquad = \left( z^2 - 1 \right)^2 [ j_+^{(2)} , j_-^{(2)} ] \, .
\end{align}
Therefore we find
\begin{align}
    \left[ \mathfrak{L}_+ , \mathfrak{L}_- \right] &\deq [ j_+^{(0)} , j_-^{(0)} ] + [ j_+^{(2)} , j_-^{(2)} ] 
    \\
    &\qquad + \frac{1}{z^2 - 1} \left( \left[ j_+^{(0)} , ( z^2 + 1 ) j_-^{(2)} + 2 z \mathfrak{J}_-^{(2)} \right] + \left[ ( z^2 + 1 ) j_+^{(2)} - 2 z \mathfrak{J}_+^{(2)} , j_-^{(0)} \right] \right) \, .
    \notag
\end{align}
The commutator terms involving $j_\pm^{(0)}$ will be precisely what we need to combine with other terms in the curvature and assemble into covariant derivatives. Indeed, we have
\begin{align}
    d \mathfrak{L} &= \partial_+ \mathcal{L}_- - \partial_- \mathfrak{L}_+ + [ \mathcal{L}_+ , \mathcal{L}_- ] \nonumber \\
    &\deq \partial_+ j_-^{(0)} - \partial_- j_+^{(0)} + [ j_+^{(0)} , j_-^{(0)} ] + [ j_+^{(2)} , j_-^{(2)} ] \nonumber \\
    &\qquad + \frac{z^2 + 1}{z^2 - 1} \left( \partial_+ j_-^{(2)} - \partial_- j_+^{(2)} + [ j_+^{(0)} , j_-^{(2)} ] - [ j_-^{(0)} , j_+^{(2)} ]  \right) \nonumber \\
    &\qquad + \frac{2 z}{z^2 - 1} \left( \partial_+ \mathfrak{J}_-^{(2)} + \partial_- \mathfrak{J}_+^{(2)} + [ j_+^{(0)} , \mathfrak{J}_-^{(2)} ] + [ j_-^{(0)} , \mathfrak{J}_+^{(2)} ] \right) \nonumber \\
    &= F_{+-}^{(0)} + [ j_+^{(2)} , j_-^{(2)} ] + \frac{z^2 + 1}{z^2 - 1} \left( D_+ j_-^{(2)} - D_- j_+^{(2)} \right) + \frac{2 z}{z^2 - 1} \left( D_+ \mathfrak{J}_-^{(2)} + D_- \mathfrak{J}_+^{(2)} \right)
    \, .
\end{align}
This is the result which we quote in equation (\ref{final_AFSSSM_lax}) of the body.

\section{Details of sSSSM-WZ Calculations}\label{app:sSSSM}

In this Appendix, we present similar computational details to those in Appendix \ref{app:SSSM}, but for the semi-symmetric space sigma model with WZ term deformed by auxiliary fields (rather than the symmetric space sigma model with auxiliary fields in the previous Appendix). In particular, we show several steps in the derivation of the equations of motion for the model (\ref{semi_plus_wz_deformed}), and in the calculation of the curvature of the Lax connection (\ref{sSSSM_WZ_lax}) for the theory.

\subsection{Equations of Motion}\label{app:sSSSM_eom}

First let us discuss the Euler-Lagrange equations for the two fields $v_{\pm}$ and $g$ in the AF-sSSSM-WZ. The structure of the auxiliary field equation of motion is essentially identical to that in the symmetric space sigma model coupled to auxiliary fields, which we explained around equation (\ref{SSSM_auxiliary_eom_app}). In particular, since only the projection $v_\alpha^{(2)}$ of the auxiliary fields onto $\mathfrak{g}_2$ appear, and in exactly the same combination as in the symmetric space sigma model, the $v_\alpha$ equation of motion is still
\begin{align}\label{sSSSM_WZ_auxiliary_eom}
    0 &\deq j_{\pm}^{(2)} + v_{\pm}^{(2)}
    \\
    &\quad + \sum_{n = 2}^{N} n \frac{\partial E}{\partial \nu_n} \tr \left( \Big( v_{\pm}^{(2)} \Big)^n \right) \Big( v_{\mp}^{(2)} \Big)^{A_1} \left( v_{\mp}^{(2)} \right)^{A_2} \ldots \left( v_{\mp}^{(2)} \right)^{A_{n-1}} T^{A_n} \tr ( T_{(A_1} T_{A_2} \ldots T_{A_{n} )} ) \, .
    \notag
\end{align}
This is convenient, since by repeating the same arguments involving the generalized Jacobi identity and the manipulations around equations (\ref{justify_one}) and (\ref{justify_two}), we conclude that all of the same commutator identities also hold in the sSSSM-WZ coupled to auxiliaries, i.e.
\begin{align}
    [ v_{\mp}^{(2)} , j_{\pm}^{(2)} ] \deq [ v_{\pm}^{(2)} , v_{\mp}^{(2)} ] \, , \quad [ \mathfrak{J}_+^{(2)} , j_-^{(2)} ] \deq [ j_+^{(2)} , \mathfrak{J}_-^{(2)} ] \, , \quad     [ \mathfrak{J}_+^{(2)} , \mathfrak{J}_-^{(2)} ] \deq [ j_+^{(2)} , j_-^{(2)} ] \, .
\end{align}
That is, these useful commutator identities are completely unaffected by the inclusion of the fermionic fields with odd grading in the model (and by the addition of the WZ term).

Next we turn to the Euler-Lagrange equation for the physical field $g$. This derivation proceeds along similar lines as in the AF-SSSM, described in Appendix \ref{app:SSSM_eom}, but with four component fields rather than two. As usual, the variation of the Maurer-Cartan form is $\delta j_{\pm} = \partial_{\pm} \epsilon + [ j_{\pm} , \epsilon ]$. We decompose $j_{\pm}$ and $\epsilon$ into components along the four subspaces $\mathfrak{g}_n$:
\begin{align}\label{four_subspace_decomp_variation}
    j_\alpha &= j_\alpha^{(0)} + j_\alpha^{(1)} + j_\alpha^{(2)} + j_\alpha^{(3)} \, , \qquad j_\alpha^{(n)} \in \mathfrak{g}_n \, , \nonumber \\
    \epsilon &= \epsilon^{(0)} + \epsilon^{(1)} + \epsilon^{(2)} + \epsilon^{(3)} \, , \qquad \; \epsilon^{(n)} \in \mathfrak{g}_n \, .
\end{align}
Then the variations of each of the projections of $j_{\pm}$ are
\begin{align}\label{sSSSM_WZ_deltaj_projected}
    \delta j_{\pm}^{(0)} &= D_{\pm} \epsilon^{(0)} + [ j_{\pm}^{(2)} , \epsilon^{(2)} ] + [ j_{\pm}^{(1)} , \epsilon^{(3)} ] + [ j_{\pm}^{(3)} , \epsilon^{(1)} ] \, , \nonumber \\
    \delta j_{\pm}^{(1)} &= D_{\pm} \epsilon^{(1)} + [ j_{\pm}^{(1)} , \epsilon^{(0)} ] + [ j_{\pm}^{(2)} , \epsilon^{(3)} ] + [ j_{\pm}^{(3)} , \epsilon^{(2)} ] \, , \nonumber \\
    \delta j_{\pm}^{(2)} &= D_{\pm} \epsilon^{(2)} + [ j_{\pm}^{(2)} , \epsilon^{(0)} ] + [ j_{\pm}^{(1)} , \epsilon^{(1)} ] + [ j_{\pm}^{(3)} , \epsilon^{(3)} ] \, , \nonumber \\
    \delta j_{\pm}^{(3)} &= D_{\pm} \epsilon^{(3)} + [ j_{\pm}^{(3)} , \epsilon^{(0)} ] + [ j_{\pm}^{(2)} , \epsilon^{(1)} ] + [ j_{\pm}^{(1)} , \epsilon^{(2)} ] \, .
\end{align}
As we mentioned in the discussion of the SSSM, in the ``physical theory'' describing the dynamics of $g$ when the auxiliaries have been integrated out, we know that $j^{(0)}$ behaves like a gauge field, and we write quantities in terms of the covariant derivative $D_{\pm} = \partial_{\pm} + [ j^{(0)}_{\pm} \, , \, \cdot \, ]$ and the field strength $F_{\alpha \beta}^{(0)} = \partial_\alpha j^{(0)}_\beta - \partial_\beta j_\alpha^{(0)} + [ j_\alpha^{(0)} , j_\beta^{(0)} ]$. From this perspective, the component $\epsilon^{(0)}$ of the variation simply enacts a gauge transformation of $j^{(0)}$ of the form
\begin{align}
    j^{(0)}_\alpha \to j_\alpha^{(0)} + \partial_\alpha \epsilon^{(0)} + [ j_\alpha^{(0)} , \epsilon^{(0)} ] \, ,
\end{align}
as well as implementing a gauge transformation on the other $\delta j_{\pm}^{(i)}$ by virtue of the properties of the covariant derivative $D_{\pm}$. Such a gauge transformation should have no effect on the physics, so long as all of our expressions are written in terms of gauge-invariant quantities. Therefore, the equation of motion associated with this component $\epsilon^{(0)}$ should be trivial when the auxiliary field equation of motion is satisfied.\footnote{This observation was also implicit in the analysis of Appendix \ref{app:SSSM}, although there we took a slightly different strategy and worked in terms of the combined variation $\epsilon$ rather than the projections $\epsilon^{(n)}$. These two approaches are complementary and illustrate that one reaches similar conclusions in different ways, which is one reason for presenting both analyses separately (even though one is a special case of the other).} We will later see that this is the case, but for now, we will explicitly keep factors proportional to $\epsilon^{(0)}$, and we will not yet impose the $v_\pm$ equation of motion when computing the variations of the action.

Now let us vary the terms appearing in the first line of (\ref{semi_plus_wz_deformed}). We consider
\begin{align}
    \delta \left( \mathrm{str} \left( j_+^{(2)} j_-^{(2)} \right) \right) &= \mathrm{str} \left( \delta j_+^{(2)} j_-^{(2)} + \delta j_-^{(2)} j_+^{(2)}  \right) \nonumber \\
    &= \mathrm{str} \Bigg( \left( D_{+} \epsilon^{(2)} + [ j_{+}^{(2)} , \epsilon^{(0)} ] + [ j_{+}^{(1)} , \epsilon^{(1)} ] + [ j_{+}^{(3)} , \epsilon^{(3)} ] \right) j_-^{(2)}  \nonumber \\
    &\qquad \qquad + \left( D_{-} \epsilon^{(2)} + [ j_{-}^{(2)} , \epsilon^{(0)} ] + [ j_{-}^{(1)} , \epsilon^{(1)} ] + [ j_{-}^{(3)} , \epsilon^{(3)} ] \right) j_+^{(2)}  \Bigg) \, ,
\end{align}
where we used the third line of (\ref{sSSSM_WZ_deltaj_projected}). When this variation is performed under the integral so that we may integrate by parts, we have
\begin{align}
    &\int_{\Sigma} d^2 \sigma \, \delta \left( \mathrm{str} \left( j_+^{(2)} j_-^{(2)} \right) \right) \nonumber \\
    &\quad = \int_{\Sigma} d^2 \sigma \, \mathrm{str} \Bigg( - \epsilon^{(2)} \left( D_+ j_-^{(2)} + D_- j_+^{(2)} \right) + \left( [ j_{+}^{(1)} , \epsilon^{(1)} ] 
    + [ j_{+}^{(3)} , \epsilon^{(3)} ] + [ j_+^{(2)} , \epsilon^{(0)} ] \right) j_-^{(2)}  \nonumber \\
    &\qquad \qquad \qquad + \left( [ j_{-}^{(1)} , \epsilon^{(1)} ] + [ j_{-}^{(3)} , \epsilon^{(3)} ] + [ j_-^{(2)} , \epsilon^{(0)} ] \right) j_+^{(2)} \Bigg) \, .
\end{align}
Simplifying the commutator terms using identities of the form (\ref{trace_commutator_identity}) then gives
\begin{align}
    &\int_{\Sigma} d^2 \sigma \, \delta \left( \mathrm{str} \left( j_+^{(2)} j_-^{(2)} \right) \right) \nonumber \\
    &\quad = \int_{\Sigma} d^2 \sigma \, \mathrm{str} \Bigg( - \epsilon^{(2)} \left( D_+ j_-^{(2)} + D_- j_+^{(2)} \right) + \epsilon^{(1)} \left( [ j_-^{(2)} , j_+^{(1)} ] + [ j_+^{(2)} , j_-^{(1)} ]  \right) \nonumber \\
    &\qquad \qquad \qquad + \epsilon^{(3)} \left( [ j_-^{(2)} , j_+^{(3)} ] + [ j_+^{(2)} , j_-^{(3)} ]  \right) + \epsilon^{(0)} \left( [ j_-^{(2)} , j_+^{(2)} ] + [ j_+^{(2)} , j_-^{(2)} ]  \right)  \Bigg) \, .
\end{align}
Note that the coefficient of the $\epsilon^{(0)}$ term vanishes identically, as expected from the intuition that this term represents a gauge transformation when the auxiliary field is on-shell. Since the variation of this term in the action involves no auxiliary fields, we therefore expect $\epsilon^{(0)}$ not to contribute, which is indeed what we find.

Next we must vary the term coupling $j_\alpha^{(2)}$ and $v_\alpha^{(2)}$. One has
\begin{align}
    \delta \left( \mathrm{str} \left( j_+^{(2)} v_-^{(2)} + j_-^{(2)} v_+^{(2)} \right) \right) &= \mathrm{str} \left( \left( \delta j_+^{(2)} \right) v_-^{(2)} + \left( \delta j_-^{(2)} \right) v_+^{(2)} \right) 
    \\
    &= \mathrm{str} \Bigg( \left( D_{+} \epsilon^{(2)} + [ j_{+}^{(2)} , \epsilon^{(0)} ] + [ j_{+}^{(1)} , \epsilon^{(1)} ] + [ j_{+}^{(3)} , \epsilon^{(3)} ] \right) v_-^{(2)} \nonumber \\
    &\quad \qquad + \left( D_{-} \epsilon^{(2)} + [ j_{-}^{(2)} , \epsilon^{(0)} ] + [ j_{-}^{(1)} , \epsilon^{(1)} ] + [ j_{-}^{(3)} , \epsilon^{(3)} ] \right) v_+^{(2)} \Bigg) \, .
    \notag
\end{align}
Under an integral, we can integrate by parts, and we simplify the commutator terms using identities like (\ref{trace_commutator_identity}) to write
\begin{align}
    &\int_{\Sigma} d^2 \sigma \, \delta \left( \mathrm{str} \left( j_+^{(2)} v_-^{(2)} + j_-^{(2)} v_+^{(2)} \right) \right) \nonumber \\
    &\quad = \int_{\Sigma} d^2 \sigma \, \mathrm{str} \Bigg( - \epsilon^{(2)} \left( D_+ v_-^{(2)} + D_- v_+^{(2)} \right) + \epsilon^{(1)} \left( [ v_-^{(2)} , j_+^{(1)} + [ v_+^{(2)} , j_-^{(1)} ] \right) \nonumber \\
    &\qquad \qquad \qquad \qquad + \epsilon^{(3)} \left( [ v_-^{(2)} , j_+^{(3)} ] + [ v_+^{(2)} , j_-^{(3)} ] \right) + \epsilon^{(0)} \left( [ v_-^{(2)} , j_+^{(2)} ] + [ v_+^{(2)} , j_-^{(2)} ] \right) \Bigg) \, .
\end{align}
In this case, the coefficient of $\epsilon^{(0)}$ does not vanish identically when all fields are off-shell, but it is equal to zero when the auxiliary field equation of motion is satisfied, as it must be.

Finally we vary the $x_0$ term. One has
\begin{align}
    \delta \left( \mathrm{str} \left( j_+^{(1)} j_-^{(3)} \!-\! j_-^{(1)} j_+^{(3)} \right) \right) &\!=\! \mathrm{str} \left( ( \delta j_+^{(1)} )  j_-^{(3)} \!+\! ( \delta j_-^{(3)} ) j_+^{(1)} \!-\! ( \delta  j_-^{(1)} )  j_+^{(3)} \!-\! ( \delta j_+^{(3)} ) j_-^{(1)} \right) \nonumber \\
    &\!=\! \mathrm{str} \Bigg( \left( D_{+} \epsilon^{(1)} \!+\! [ j_{+}^{(1)} , \epsilon^{(0)} ] \!+\! [ j_{+}^{(2)} , \epsilon^{(3)} ] \!+\! [ j_{+}^{(3)} , \epsilon^{(2)} ] \right)  j_-^{(3)} \nonumber \\
    &\qquad \!+\! \left( D_{-} \epsilon^{(3)} \!+\! [ j_{-}^{(3)} , \epsilon^{(0)} ]  + [ j_{-}^{(2)} , \epsilon^{(1)} ] \!+\! [ j_{-}^{(1)} , \epsilon^{(2)} ] \right)  j_+^{(1)} \nonumber \\
    &\qquad \!-\! \left( D_{-} \epsilon^{(1)} \!+\! [ j_{-}^{(1)} , \epsilon^{(0)} ] \!+\! [ j_{-}^{(2)} , \epsilon^{(3)} ] \!+\! [ j_{-}^{(3)} , \epsilon^{(2)} ] \right)   j_+^{(3)} \nonumber \\
    &\qquad \!-\! \left( D_{+} \epsilon^{(3)} \!+\! [ j_{+}^{(3)} , \epsilon^{(0)} ]  \!+\! [ j_{+}^{(2)} , \epsilon^{(1)} ] \!+\! [ j_{+}^{(1)} , \epsilon^{(2)} ] \right) j_-^{(1)} \Bigg) \, ,
\end{align}
where we have used the variations (\ref{sSSSM_WZ_deltaj_projected}). When this variation is performed under an integral, we may integrate by parts to write
\begin{align}
    \int_{\Sigma} &d^2 \sigma \, \delta x_0 \nonumber \\
    &= \frac{1}{4} \int_{\Sigma} d^2 \sigma \, \mathrm{str} \Bigg( \epsilon^{(1)} \left( D_- j_3^{(+)} - D_+ j_-^{(3)} \right) + \epsilon^{(3)} \left( D_+ j_-^{(1)} - D_- j_+^{(1)} \right)
    \\
    &\quad  + \left( [ j_+^{(1)} , \epsilon^{(0)} ] + [ j_{+}^{(2)} , \epsilon^{(3)} ]  + [ j_{+}^{(3)} , \epsilon^{(2)} ] \right)  j_-^{(3)} + \left( [ j_-^{(3)} , \epsilon^{(0)} ] + [ j_{-}^{(2)} , \epsilon^{(1)} ] + [ j_{-}^{(1)} , \epsilon^{(2)} ] \right)  j_+^{(1)} \nonumber \\
    &\quad - \left( [ j_-^{(1)} , \epsilon^{(0)} ] + [ j_{-}^{(2)} , \epsilon^{(3)} ] + [ j_{-}^{(3)} , \epsilon^{(2)} ] \right)   j_+^{(3)} - \left( [ j_+^{(3)} , \epsilon^{(0)} ] +  [ j_{+}^{(2)} , \epsilon^{(1)} ] + [ j_{+}^{(1)} , \epsilon^{(2)} ] \right) j_-^{(1)} \Bigg) \, .
    \notag
\end{align}
Again using identities like (\ref{trace_commutator_identity}), this is 
\begin{align}
    \int_{\Sigma} &d^2 \sigma \, \delta x_0 \nonumber \\
    &\!=\! \frac{1}{4} \!\int_{\Sigma}\! \!d^2\!\sigma \, \mathrm{str} \!\Bigg(\! \!\epsilon^{(1)}\!
    \left( D_- j_3^{(+)} \!-\! D_+ j_-^{(3)} \right) \!+\! \epsilon^{(3)} \left( D_+ j_-^{(1)} \!-\! D_- j_+^{(1)} \right) \!+\! \epsilon^{(3)} \left( [ j_-^{(2)} , j_+^{(3)} ] \!-\! [ j_+^{(2)} , j_-^{(3)} ] \right)\nonumber \\
    &\qquad \qquad \,\,\, + \epsilon^{(1)} \left(  [ j_+^{(2)} , j_-^{(1)} ] \!+\! [ j_+^{(1)}  , j_-^{(2)} ]  \right) \!-\! 2 \epsilon^{(2)} \left( [ j_+^{(3)} , j_-^{(3)} ] \!+\! [ j_-^{(1)} , j_+^{(1)} ] \right) \nonumber \\
    &\qquad  \qquad \,\,\, + \epsilon^{(0)} \left( [ j_-^{(3)} , j_+^{(1)} ] \!+\! [ j_+^{(1)} , j_-^{(3)} ] \!-\! [ j_+^{(3)} , j_-^{(1)} ] \!-\! [ j_-^{(1)} , j_+^{(3} ]  \right) \Bigg) \, .
\end{align}
Again we note that the coefficient of the $\epsilon^{(0)}$ term vanishes identically, as it must since this term is independent of auxiliary fields.

Finally, let us study the contribution to the equations of motion from the Wess-Zumino term, since this will introduce an important subtlety. Much like the steps leading up to equation (\ref{final_delta_SWZ}), the variation of the sSSSM Wess-Zumino term leads to a total derivative on $\mathcal{M}_3$, plus terms that vanish upon using the Maurer-Cartan identity and Jacobi identity. These steps are the supersymmetric generalization of those that we reviewed in Section \ref{sec:AFSM_WZ} so we will not present all of the intermediate steps, although some such details are discussed in \cite{Cagnazzo:2012se} and in Appendix A.3 of \cite{Borsato:2022tmu}. The conclusion of this calculation is that
\begin{equation}
    \delta S_{\text{WZ}}^{\text{sSSSM}} = \kay \int_{\mathcal{M}_3} \epsilon^{ijk} \Bigl( \partial_{i}B_{jk} + Z_{ijk} \Bigr)
    \, ,
\end{equation}
with 
\begin{equation}
    B_{jk}= \mathrm{str} \left( \epsilon^{(2)} ( [j^{(2)}_j ,j^{(2)}_k ] +[j^{(1)}_j ,j^{(3)}_k ] ) +\epsilon^{(1)}[j^{(3)}_j ,j^{(2)}_k ]+\epsilon^{(3)}[j^{(1)}_j ,j^{(2)}_k ] \right)
    \, ,
\end{equation}
and
\begin{align}
    Z_{ijk}=- & \mathrm{str} 
    \Bigl(  \epsilon^{(1)}\bigl( D_{i}[j_{j}^{(3)},j_{k}^{(2)}]\!+ \![j_{i}^{(1)},[j_{j}^{(3)},j_{k}^{(1)}]]\!+\![j_{i}^{(2)},[j_{j}^{(2)},j_{k}^{(1)}]]\!+\![j_{i}^{(1)},[j_{j}^{(2)},j_{k}^{(2)}]]\bigr)
    \notag \\
    & \,\,\,\, +
    \epsilon^{(2)}\bigl( D_{i}[j_{j}^{(3)},j_{k}^{(1)}]\!+ \!D_{i}[j_{j}^{(2)},j_{k}^{(2)}]\!+\![j_{i}^{(1)},[j_{j}^{(2)},j_{k}^{(1)}]]\!+\![j_{i}^{(3)},[j_{j}^{(3)},j_{k}^{(2)}]]\bigr)
    \\
    & \,\,\,\, +
    \epsilon^{(3)}\bigl( D_{i}[j_{j}^{(2)},j_{k}^{(1)}]\!+ \![j_{i}^{(2)},[j_{j}^{(3)},j_{k}^{(2)}]]\!+\![j_{i}^{(3)},[j_{j}^{(3)},j_{k}^{(1)}]]\!+\![j_{i}^{(3)},[j_{j}^{(2)},j_{k}^{(2)}]]\bigr)
    \Bigr)
    \, .
    \notag
\end{align}
In particular $Z_{ijk}$ vanishes identically as a consequence of the Maurer-Cartan identity and the Jacobi identity. As a result, the net contribution from the variation of the WZ term localizes to an integral over $\Sigma$, namely
\begin{align}\label{delta_S_WZ}
    \delta S_{\text{WZ}}^{\text{sSSSM}} &= \frac{\kay}{2} \int_{\Sigma} \mathrm{str} \Bigg( \epsilon^{(2)} \left( 2 [ j^{(2)}_+ ,j^{(2)}_- ] + [j^{(1)}_+ ,j^{(3)}_- ] - [j^{(1)}_- ,j^{(3)}_+ ] \right) + \epsilon^{(1)} \left( [ j^{(3)}_+ ,j^{(2)}_- 
    ] - [ j^{(3)}_- ,j^{(2)}_+ 
    ] \right) \nonumber \\
    &\qquad \qquad \qquad \qquad + \epsilon^{(3)} \left( [ j^{(1)}_+ ,j^{(2)}_- ] - [ j^{(1)}_- ,j^{(2)}_+ ] \right) \Bigg) \, .
\end{align}
Note that the overall sign in the final line of (\ref{delta_S_WZ}) depends on a sign convention for the orientation of $\mathcal{M}_3$; here we have made a particular choice, but one can always reverse this convention by taking $\kay \to - \kay$.
Putting together the pieces above and collecting the terms proportional to each of $\epsilon^{(1)}$, $\epsilon^{(2)}$, $\epsilon^{(3)}$, we have
\begin{align}\label{AFsSSSM_WZ_eom_intermediate}
    \delta S = \int &d^2 \sigma \, \Bigg( \epsilon^{(1)} \Big( \frac{\hay}{2} \left( [ j_-^{(2)} , j_+^{(1)} ] + [ j_+^{(2)} , j_-^{(1)} ]  \right) + \hay \left( [ v_-^{(2)} , j_+^{(1)} ] + [ v_+^{(2)} , j_-^{(1)} ] \right) \nonumber \\
    &\quad + \frac{\ell \hay}{4} \left( D_- j_3^{(+)} - D_+ j_-^{(3)} + [ j_+^{(2)} , j_-^{(1)} ] - [ j_-^{(2)} , j_+^{(1)} ] \right) + \frac{\kay}{2} \left( [j^{(3)}_+ ,j^{(2)}_- ] - [ j^{(3)}_- , j^{(2)}_+ ] \right)  \Big) \nonumber \\
    &+ \epsilon^{(2)} \Big( - \frac{\hay}{2} \left( D_+ j_-^{(2)} + D_- j_+^{(2)} \right) - \hay \left( D_+ v_-^{(2)} + D_- v_+^{(2)} \right) \nonumber \\
    &\quad  - \frac{\ell \hay}{2} \left( [ j_+^{(3)} , j_-^{(3)} ] + [ j_-^{(1)} , j_+^{(1)} ] \right) + \frac{\kay}{2} \left( 2 [j^{(2)}_+ ,j^{(2)}_- ] +[j^{(1)}_+ ,j^{(3)}_- ] - [ j^{(1)}_- , j^{(3)}_+ ] \right)  \Big) \nonumber \\
    &+ \epsilon^{(3)} \Big( \frac{\hay}{2} \left( [ j_-^{(2)} , j_+^{(3)} ] + [ j_+^{(2)} , j_-^{(3)} ]  \right) + \hay \left( [ v_-^{(2)} , j_+^{(3)} ] + [ v_+^{(2)} , j_-^{(3)} ] \right)  \nonumber \\
    &\quad + \frac{\ell \hay}{4} \left( D_+ j_-^{(1)} - D_- j_+^{(1)} + [ j_-^{(2)} , j_+^{(3)} ]  -  [ j_+^{(2)} , j_-^{(3)} ] \right) + \frac{\kay}{2} \left( [j^{(1)}_+ ,j^{(2)}_- ] - [ j^{(1)}_- , j^{(2)}_+ ] \right) \Big) \nonumber \\
    &\quad + \epsilon^{(0)} \left( [ v_-^{(2)} , j_+^{(2)} ] + [ v_+^{(2)} , j_-^{(2)} ] \right) \Bigg) \, .
\end{align}
Note that, as we mentioned in Section \ref{sec:sSSSM}, the presence of the Wess-Zumino term breaks the $\mathbb{Z}_4$ grading, so we can no longer decompose a given equation into its projections onto the four subspaces $\mathfrak{g}_n$. However, what we \emph{can} still do is decompose the equations of motion into components along the three variations $\epsilon^{(1)}$, $\epsilon^{(2)}$, and $\epsilon^{(3)}$, since we are free to perform each of these variations separately, and the action must be stationary with respect to each of the three fluctuations.  Therefore, setting each of the independent variations in (\ref{AFsSSSM_WZ_eom_intermediate}) to zero, and dividing through to clear constants, we arrive at the equations of motion
\begin{align}\label{unsimplified_sSSSM_eom}
    0 &= [ v_-^{(2)} , j_+^{(2)} ] + [ v_+^{(2)} , j_-^{(2)} ] \, , \nonumber \\
    0 &= 2 \left( [ j_-^{(2)} + 2 v_-^{(2)} , j_+^{(1)} ] + [ j_+^{(2)} + 2 v_+^{(2)} , j_-^{(1)} ]  \right) + \ell \left( D_- j_3^{(+)} - D_+ j_-^{(3)} + [ j_+^{(2)} , j_-^{(1)} ] - [ j_-^{(2)} , j_+^{(1)} ] \right)  \nonumber \\
    &\qquad + \frac{2 \kay}{\hay} \left( [j^{(3)}_+ ,j^{(2)}_- ] - [ j^{(3)}_- , j^{(2)}_+ ] \right) \, , \nonumber \\
    0 &=   D_+ \left( j_-^{(2)} + 2 v_-^{(2)} \right) + D_- \left( j_+^{(2)} + 2 v_+^{(2)} \right)  + \ell \left( [ j_+^{(3)} , j_-^{(3)} ] - [ j_+^{(1)} , j_-^{(1)} ] \right) \nonumber \\
    &\qquad - \frac{\kay}{\hay} ( 2 [j^{(2)}_+ ,j^{(2)}_- ]  + [j^{(1)}_+ ,j^{(3)}_- ] - [ j^{(1)}_- , j^{(3)}_+ ] )  \, , \nonumber \\
    0 &= 2 \left( [ j_-^{(2)} + 2 v_-^{(2)} , j_+^{(3)} ] + [ j_+^{(2)} + 2 v_+^{(2)} , j_-^{(3)} ]  \right) + \ell \left( D_+ j_-^{(1)} - D_- j_+^{(1)} - \left( [j_+^{(3)} ,  j_-^{(2)} ]  + [ j_+^{(2)} , j_-^{(3)} ] \right) \right) \nonumber \\
    &\qquad + \frac{2 \kay}{\hay} \left( [j^{(1)}_+ ,j^{(2)}_- ] - [ j^{(1)}_- , j^{(2)}_+ ] \right)  \, .
\end{align}
First note that, when the auxiliary field equation of motion is satisfied, the first line of (\ref{unsimplified_sSSSM_eom}) -- which corresponds to the $\epsilon^{(0)}$ variation -- is automatically satisfied, since
\begin{align}
    [ v_-^{(2)} , j_+^{(2)} ] + [ v_+^{(2)} , j_-^{(2)} ] \deq 0 \, .
\end{align}
We will therefore use dot-equality $\deq$ rather than equality $=$ in what follows, to remind us that we have already assumed that the auxiliary field equation of motion is satisfied in order to ignore the first line of (\ref{unsimplified_sSSSM_eom}). Focusing on the equations arising from the remaining three variations in terms of the field
\begin{align}
    \mathfrak{J}_{\pm}^{(2)} = - \left( j_{\pm}^{(2)} + 2 v_{\pm}^{(2)} \right) \, ,
\end{align}
these three non-trivial equations of motion can be written as
\begin{align}\label{sSSSM_eom_no_MC}
    0 &\deq - 2 \left( [ \mathfrak{J}_{-}^{(2)} , j_+^{(1)} ] + [ \mathfrak{J}_{+}^{(2)} , j_-^{(1)} ]  \right) + \ell \left( D_- j_3^{(+)} - D_+ j_-^{(3)} + [ j_+^{(2)} , j_-^{(1)} ] - [ j_-^{(2)} , j_+^{(1)} ] \right) \nonumber \\
    &\qquad + \frac{2 \kay}{\hay} \left( [j^{(3)}_+ ,j^{(2)}_- ] - [ j^{(3)}_- , j^{(2)}_+ ] \right) \, , \nonumber \\
    0 &\deq D_+ \mathfrak{J}_{-}^{(2)} + D_- \mathfrak{J}_{+}^{(2)} - \ell \left( [ j_+^{(3)} , j_-^{(3)} ] - [ j_+^{(1)} , j_-^{(1)} ] \right) + \frac{\kay}{\hay} \left( 2 [j^{(2)}_+ ,j^{(2)}_- ]  + [j^{(1)}_+ ,j^{(3)}_- ] - [ j^{(1)}_- , j^{(3)}_+ ] \right) \, , \nonumber \\
    0 &\deq - 2 \left( [ \mathfrak{J}_{-}^{(2)} , j_+^{(3)} ] + [ \mathfrak{J}_{+}^{(2)}  , j_-^{(3)} ]  \right) + \ell \left( D_+ j_-^{(1)} - D_- j_+^{(1)} + [ j_-^{(2)} , j_+^{(3)} ]  -  [ j_+^{(2)} , j_-^{(3)} ] \right) \nonumber \\
    &\qquad + \frac{2 \kay}{\hay} \left( [j^{(1)}_+ ,j^{(2)}_- ] -  [ j^{(1)}_- , j^{(2)}_+ ] \right) \, .
\end{align}
Our ultimate goal is to show that these equations of motion are equivalent to the flatness of a Lax connection for any $z \in \mathbb{C}$, assuming that the Maurer-Cartan identity holds. Since we will eventually need to use Maurer-Cartan anyway, it is convenient to simplify the equations (\ref{sSSSM_eom_no_MC}) using the Maurer-Cartan identity; if we can show that these simplified equations of motion are equivalent to flatness of a Lax, this still establishes weak integrability.

Simplifying the first and third lines of (\ref{sSSSM_eom_no_MC}) using (\ref{sSSSM_four_maurer_cartan_nicer}) gives
\begin{align}\label{sSSSM_eom_with_MC_app}
    0 &\!\deq\! [ \mathfrak{J}^{(2)}_+ , j^{(1)}_- ] \!+\! [ \mathfrak{J}^{(2)}_- , j^{(1)}_+ ] \!-\!  \ell \left( [ j^{(2)}_+ , j^{(1)}_- ] \!-\! [ j^{(2)}_- , j^{(1)}_+ ] \right) \!-\! \frac{\kay}{\hay} \left(  [ j^{(3)}_+  , j^{(2)}_- ] \!-\! [ j^{(3)}_- , j^{(2)}_+  ] \right) \, , \nonumber \\
    0 &\!\deq\! D_+ \mathfrak{J}^{(2)}_- \!+\! D_- \mathfrak{J}^{(2)}_+  \!-\! \ell \left( [ j_+^{(3)} , j_-^{(3)} ] \!-\! [ j_+^{(1)} , j_-^{(1)} ] \right) \!+\! \frac{\kay}{\hay} \left( 2 [ j_+^{(2)} , j_-^{(2)} ]  \!+\! [ j_+^{(1)} , j_-^{(3)} ] \!-\! [ j_-^{(1)} , j_+^{(3)} ] \right) \, , \nonumber \\
    0 &\!\deq\! [ \mathfrak{J}^{(2)}_+ , j^{(3)}_- ] \!+\! [ \mathfrak{J}^{(2)}_- , j^{(3)}_+ ] \!+\! \ell \left( [ j^{(2)}_+ , j^{(3)}_- ] \!-\! [ j^{(2)}_- , j^{(3)}_+ ] \right) \!-\! \frac{\kay}{\hay} \left( [ j^{(1)}_+ , j^{(2)}_- ] \!-\! [ j^{(1)}_- , j^{(2)}_+  ] \right) \, .
\end{align}
This is the form of the equations of motion which we use in the body of the manuscript.

\subsection{Curvature of Lax Connection}\label{app:sSSSM_curvature}

In this section we will analyze the flatness condition for the Lax connection (\ref{sSSSM_WZ_lax}) of the auxiliary field semi-symmetric space sigma model with Wess-Zumino term, or AF-sSSSM-WZ. As we noted in the body of this article, the only relations which we will need in order to demonstrate (weak) classical integrability of the theory are
\begin{align}\label{sSSSM_comm_appendix}
    [ \mathfrak{J}^{(2)}_+ , j_-^{(2)}] \deq [ j_+^{(2)} , \mathfrak{J}_-^{(2)} ] \, , \qquad [ \mathfrak{J}_+^{(2)} , \mathfrak{J}_-^{(2)} ] \deq [ j_+^{(2)} , j_-^{(2)} ] \, , 
\end{align}
which hold when the auxiliary field equation of motion is satisfied. These are the same relations which were used to analyze the sSSSM deformed by a combination of the $\TT$ and root-$\TT$ flows in \cite{Borsato:2022tmu}. As a result, the manipultions in this Appendix are very similar to those in this earlier work. However, let us emphasize that the present analysis is more general, since the coupling to auxiliary fields can accommodate deformations by \emph{general} functions of the stress tensor (rather than only $\TT$ and root-$\TT$), in addition to interactions involving higher-spin combinations of auxiliary fields, which are related (at least at leading order) to deformations by higher-spin conserved currents of Smirnov-Zamolodchikov type.

We begin by computing the commutator of $\mathfrak{L}_+$ with $\mathfrak{L}_-$, which is the third term in the curvature of the Lax connection $d_\mathfrak{L} \mathfrak{L} = \partial_+ \mathfrak{L}_- - \partial_- \mathfrak{L}_+ + [ \mathfrak{L}_+ , \mathfrak{L}_-]$. One finds
\allowdisplaybreaks
\begin{align}\label{AFsSSSMWZ_lax_lax_commutator}
    [ \mathfrak{L}_+ , \mathfrak{L}_- ] &= [ j^{(0)}_+ , j^{(0)}_- ] + \left( \ell \frac{z^2 + 1}{z^2 - 1} \right)^2 [ j_+^{(2)} , j_-^{(2)} ] - \left( \frac{\kay}{\hay} - \frac{2 \ell z}{z^2 - 1 } \right)^2 [ \mathfrak{J}_+^{(2)} , \mathfrak{J}_-^{(2)} ] \nonumber \\
    & + \left( z + \frac{\ell}{1 - \frac{\kay}{\hay} } \right)^2 \left( \frac{ \ell \left( 1 - \frac{\kay}{\hay} \right) }{z^2 - 1 } \right) [ j_+^{(1)} , j_-^{(1)} ] + \left( z - \frac{\ell}{1 + \frac{\kay}{\hay} } \right)^2 \left( \frac{ \ell \left(  1 + \frac{\kay}{\hay} \right) }{ z^2 - 1 } \right) [ j_+^{(3)} , j_-^{(3)} ]  \nonumber \\
    & + \ell \frac{z^2 + 1}{z^2 - 1} \left( [ j^{(0)}_+ , j^{(2)}_- ] + [ j^{(2)}_+ , j^{(0)}_- ] \right) + \left( \frac{\kay}{\hay} - \frac{2 \ell z}{z^2 - 1 } \right) \left( [ \mathfrak{J}_+^{(2)} , j^{(0)}_- ] - [ j^{(0)}_+ , \mathfrak{J}_-^{(2)} ] \right) \nonumber \\
    & + \left( z + \frac{\ell}{1 - \frac{\kay}{\hay} } \right) \sqrt{ \frac{ \ell \left( 1 - \frac{\kay}{\hay} \right) }{z^2 - 1 } } \left( [ j_+^{(0)} , j_-^{(1)} ] + [ j_+^{(1)} , j_-^{(0)} ] \right) \\
    & + \left( z - \frac{\ell}{1 + \frac{\kay}{\hay} } \right) \sqrt{ \frac{ \ell \left( 1 + \frac{\kay}{\hay} \right) }{ z^2 - 1 } } \left( [ j_+^{(0)} , j_-^{(3)}] + [ j_+^{(3)} , j_-^{(0)} ] \right) \nonumber \\
    & + \ell \frac{z^2 + 1}{z^2 - 1} \left( \frac{\kay}{\hay} - \frac{2 \ell z}{z^2 - 1 } \right)  \left( [ \mathfrak{J}_+^{(2)} , j_-^{(2)} ] - [ j_+^{(2)} , \mathfrak{J}_-^{(2)} ]  \right) \nonumber \\
    & + \ell \frac{z^2 + 1}{z^2 - 1} \left( z + \frac{\ell}{1 - \frac{\kay}{\hay} } \right) \sqrt{ \frac{ \ell \left( 1 - \frac{\kay}{\hay} \right) }{z^2 - 1 } }  \left( [ j_+^{(2)} , j_-^{(1)} ] + [ j_+^{(1)} , j_-^{(2)} ] \right) \nonumber \\
    & + \ell \frac{z^2 + 1}{z^2 - 1} \left( z - \frac{\ell}{1 + \frac{\kay}{\hay} } \right) \sqrt{ \frac{ \ell \left( 1 + \frac{\kay}{\hay} \right) }{ z^2 - 1 } }  \left( [ j_+^{(2)} , j_-^{(3)} ] + [ j_+^{(3)} , j_-^{(2)} ] \right) \nonumber \\
    & + \left( \frac{\kay}{\hay} - \frac{2 \ell z}{z^2 - 1 } \right) \left( z + \frac{\ell}{1 - \frac{\kay}{\hay} } \right) \sqrt{ \frac{ \ell \left( 1 - \frac{\kay}{\hay} \right) }{z^2 - 1 } }  \left( [ \mathfrak{J}_+^{(2)} , j_-^{(1)} ] - [ j_+^{(1)} , \mathfrak{J}_-^{(2)} ] \right) \nonumber \\
    & + \left( \frac{\kay}{\hay} - \frac{2 \ell z}{z^2 - 1 } \right) \left( z - \frac{\ell}{1 + \frac{\kay}{\hay} } \right) \sqrt{ \frac{ \ell \left( 1 + \frac{\kay}{\hay} \right) }{ z^2 - 1 } } \left( [ \mathfrak{J}_+^{(2)} , j_-^{(3)} ] - [ j_+^{(3)} , \mathfrak{J}_-^{(2)} ] \right) \nonumber \\
    & + \left( z + \frac{\ell}{1 - \frac{\kay}{\hay} } \right) \sqrt{ \frac{ \ell \left( 1 - \frac{\kay}{\hay} \right) }{z^2 - 1 } } \left( z - \frac{\ell}{1 + \frac{\kay}{\hay} } \right) \sqrt{ \frac{ \ell \left( 1 + \frac{\kay}{\hay} \right) }{ z^2 - 1 } } \left( [ j_+^{(1)} , j_-^{(3)} ] + [ j_+^{(3)} , j_-^{(1)} ] \right) \, .
    \notag
 \end{align}
Let us remind the reader that we cannot separately project this commutator onto each of the four subspaces $\mathfrak{g}_n$ for $n = 0, 1, 2, 3$, since we eventually wish to show that the flatness of this Lax connection is equivalent to the equations of motion, and these equations of motion explicitly violate the $\mathbb{Z}_4$ grading due to the presence of the Wess-Zumino term. However, the Euler-Lagrange equations still respect the $\mathbb{Z}_2$ grading into bosons and fermions. Therefore, writing $\mathfrak{g}_B = \mathfrak{g}_0 \oplus \mathfrak{g}_2$ and $\mathfrak{g}_F = \mathfrak{g}_1 \oplus \mathfrak{g}_3$, we may project onto these two subspaces. In the bosonic part, we then simplify the commutators involving $j_\pm^{(2)}$ and $\mathfrak{J}_{\pm}^{(2)}$ using (\ref{sSSSM_comm_appendix}) to find
\begin{align}\label{commutator_bos_ferm_two}
    [ \mathfrak{L}_+ , \mathfrak{L}_- ] \big\vert_{\mathfrak{g}_B} &\deq [ j^{(0)}_+ , j^{(0)}_- ] \!+\! \left( \ell^2  \left( \frac{z^2 \!+\! 1}{z^2 \!-\! 1} \right)^2 \!-\! \left( \frac{\kay}{\hay} \!-\! \frac{2 \ell z}{z^2 \!-\! 1 } \right)^2 \right) [ j_+^{(2)} , j_-^{(2)} ] 
    \\
    & + \left( z \!+\! \frac{\ell}{1 \!-\! \frac{\kay}{\hay} } \right)^2 \left( \frac{ \ell \left( 1 \!-\! \frac{\kay}{\hay} \right) }{z^2 \!-\! 1 } \right) [ j_+^{(1)} , j_-^{(1)} ] \!+\! \left( z \!-\! \frac{\ell}{1 \!+\! \frac{\kay}{\hay} } \right)^2 \left( \frac{ \ell \left(  1 \!+\! \frac{\kay}{\hay} \right) }{ z^2 \!-\! 1 } \right) [ j_+^{(3)} , j_-^{(3)} ]  \nonumber \\
    & + \ell \frac{z^2 \!+\! 1}{z^2 \!-\! 1} \left( [ j^{(0)}_+ , j^{(2)}_- ] \!+\! [ j^{(2)}_+ , j^{(0)}_- ] \right) \!+\! \left( \frac{\kay}{\hay} \!-\! \frac{2 \ell z}{z^2 \!-\! 1 } \right) \left( [ \mathfrak{J}_+^{(2)} , j^{(0)}_- ] \!-\! [ j^{(0)}_+ , \mathfrak{J}_-^{(2)} ] \right) \nonumber \\
    & - \left( z \!+\! \frac{\ell}{1 \!-\! \frac{\kay}{\hay} } \right) \sqrt{ \frac{ \ell \left( 1 \!-\! \frac{\kay}{\hay} \right) }{z^2 \!-\! 1 } } \left( z \!-\! \frac{\ell}{1 \!+\! \frac{\kay}{\hay} } \right) \sqrt{ \frac{ \ell \left( 1 \!+\! \frac{\kay}{\hay} \right) }{ z^2 \!-\! 1 } } \left( F_{+-}^{(0)} \!+\! [ j_+^{(2)} , j_-^{(2)} ]  \right) \, , \nonumber \\
    [ \mathfrak{L}_+ , \mathfrak{L}_- ] \big\vert_{\mathfrak{g}_F} &= \left( z \!+\! \frac{\ell}{1 - \frac{\kay}{\hay} } \right) \sqrt{ \frac{ \ell \left( 1 \!-\! \frac{\kay}{\hay} \right) }{z^2 \!-\! 1 } } \left( [ j_+^{(0)} , j_-^{(1)} ] \!+\! [ j_+^{(1)} , j_-^{(0)} ] \right) \nonumber \\
    & + \left( z \!-\! \frac{\ell}{1 \!+\! \frac{\kay}{\hay} } \right) \sqrt{ \frac{ \ell \left( 1 \!+\! \frac{\kay}{\hay} \right) }{ z^2 \!-\! 1 } } \left( [ j_+^{(0)} , j_-^{(3)}] \!+\! [ j_+^{(3)} , j_-^{(0)} ] \right) \nonumber \\
    & + \ell \frac{z^2 \!+\! 1}{z^2 \!-\! 1} \left( z \!+\! \frac{\ell}{1 \!-\! \frac{\kay}{\hay} } \right) \sqrt{ \frac{ \ell \left( 1 \!-\! \frac{\kay}{\hay} \right) }{z^2 \!-\! 1 } }  \left( [ j_+^{(1)} , j_-^{(2)} ] \!+\! [ j_+^{(2)} , j_-^{(1)} ] \right) \nonumber \\
    & + \ell \frac{z^2 \!+\! 1}{z^2 \!-\! 1} \left( z \!-\! \frac{\ell}{1 \!+\! \frac{\kay}{\hay} } \right) \sqrt{ \frac{ \ell \left( 1 \!+\! \frac{\kay}{\hay} \right) }{ z^2 \!-\! 1 } }  \left( [ j_+^{(2)} , j_-^{(3)} ] \!+\! [ j_+^{(3)} , j_-^{(2)} ]  \right) \nonumber \\
    & + \left( \frac{\kay}{\hay} \!-\! \frac{2 \ell z}{z^2 \!-\! 1 } \right) \left( z \!+\! \frac{\ell}{1 \!-\! \frac{\kay}{\hay} } \right) \sqrt{ \frac{ \ell \left( 1 \!-\! \frac{\kay}{\hay} \right) }{z^2 \!-\! 1 } }  \left( [ \mathfrak{J}_+^{(2)} , j_-^{(1)} ] \!-\! [ j_+^{(1)} , \mathfrak{J}_-^{(2)} ] \right) \nonumber \\
    & + \left( \frac{\kay}{\hay} \!-\! \frac{2 \ell z}{z^2 \!-\! 1 } \right) \left( z \!-\! \frac{\ell}{1 \!+\! \frac{\kay}{\hay} } \right) \sqrt{ \frac{ \ell \left( 1 \!+\! \frac{\kay}{\hay} \right) }{ z^2 \!-\! 1 } } \left( [ \mathfrak{J}_+^{(2)} , j_-^{(3)} ] \!-\! [ j_+^{(3)} , \mathfrak{J}_-^{(2)} ] \right) \, .
 \end{align}
Note that the auxiliary field equation of motion has been used in simplifying the bosonic projection but not the fermionic contribution.

On the other hand, the derivatives of the Lax connection projected onto the bosonic and fermionic subspaces are
\begin{equation}\label{deriv_bos_ferm}
\begin{aligned}
    \left( \partial_+ \mathfrak{L}_- - \partial_- \mathfrak{L}_+ \right) \big\vert_{\mathfrak{g}_B} &= \partial_+ j_-^{(0)} - \partial_- j_+^{(0)} + \ell \frac{z^2 + 1}{z^2 - 1} \left( \partial_+ j_-^{(2)} - \partial_- j_+^{(2)} \right) \\
    &\qquad - \left( \frac{\kay}{\hay} - \frac{2 \ell z}{z^2 - 1 } \right) \left( \partial_- \mathfrak{J}_+^{(2)} + \partial_+ \mathfrak{J}_-^{(2)} \right) \, , 
    \\
    \left( \partial_+ \mathfrak{L}_- -  \partial_- \mathfrak{L}_+ \right) \big\vert_{\mathfrak{g}_F} &= \left( z + \frac{\ell}{1 - \frac{\kay}{\hay} } \right) \sqrt{ \frac{ \ell \left( 1 - \frac{\kay}{\hay} \right) }{z^2 - 1 } } \left( \partial_+ j_{-}^{(1)} - \partial_- j_+^{(1)} \right) \\
    &\quad + \left( z - \frac{\ell}{1 + \frac{\kay}{\hay} } \right) \sqrt{ \frac{ \ell \left( 1 + \frac{\kay}{\hay} \right) }{ z^2 - 1 } } \left( \partial_+ j_{-}^{(3)} - \partial_- j_+^{(3)} \right) \, .
\end{aligned}
\end{equation}
Let us now combine equations (\ref{commutator_bos_ferm_two}) and (\ref{deriv_bos_ferm}) to obtain the projections of the curvature of the Lax onto the bosonic and fermionic subspaces. The bosonic part is
\begin{align}\label{gbos_intermediate_no_mc}
    &\left( d_{\mathfrak{L}} \mathfrak{L} \right) \big\vert_{\mathfrak{g}_B} \nonumber \\
    &\quad \deq F_{+-}^{(0)} + \ell \frac{z^2 + 1}{z^2 - 1} \left( \partial_+ j_-^{(2)} - \partial_- j_+^{(2)} \right) -  \left( \frac{\kay}{\hay} - \frac{2 \ell z}{z^2 - 1 } \right) \left( \partial_- \mathfrak{J}_+^{(2)} + \partial_+ \mathfrak{J}_-^{(2)} \right) \nonumber \\
    &\qquad + \left( \ell^2  \left( \frac{z^2 + 1}{z^2 - 1} \right)^2 - \left( \frac{\kay}{\hay} - \frac{2 \ell z}{z^2 - 1 } \right)^2 \right) [ j_+^{(2)} , j_-^{(2)} ] \nonumber \\
    &\qquad + \left( z + \frac{\ell}{1 - \frac{\kay}{\hay} } \right)^2 \left( \frac{ \ell \left( 1 - \frac{\kay}{\hay} \right) }{z^2 - 1 } \right) [ j_+^{(1)} , j_-^{(1)} ] + \left( z - \frac{\ell}{1 + \frac{\kay}{\hay} } \right)^2 \left( \frac{ \ell \left(  1 + \frac{\kay}{\hay} \right) }{ z^2 - 1 } \right) [ j_+^{(3)} , j_-^{(3)} ]  \nonumber \\
    &\qquad + \ell \frac{z^2 + 1}{z^2 - 1} \left( [ j^{(0)}_+ , j^{(2)}_- ] + [ j^{(2)}_+ , j^{(0)}_- ] \right) + \left( \frac{\kay}{\hay} - \frac{2 \ell z}{z^2 - 1 } \right) \left( [ \mathfrak{J}_+^{(2)} , j^{(0)}_- ] - [ j^{(0)}_+ , \mathfrak{J}_-^{(2)} ] \right) \nonumber \\
    &\qquad - \left( z + \frac{\ell}{1 - \frac{\kay}{\hay} } \right) \sqrt{ \frac{ \ell \left( 1 - \frac{\kay}{\hay} \right) }{z^2 - 1 } } \left( z - \frac{\ell}{1 + \frac{\kay}{\hay} } \right) \sqrt{ \frac{ \ell \left( 1 + \frac{\kay}{\hay} \right) }{ z^2 - 1 } } \left( F_{+-}^{(0)} + [ j_+^{(2)} , j_-^{(2)} ]  \right) \, ,
\end{align}
where we combined the terms $\partial_+ j_-^{(0)} - \partial_- j_+^{(0)} + [ j^{(0)}_+ , j^{(0)}_- ] $ into $F_{+-}^{(0)}$ using the definition (\ref{SSSM_field_strength}).

Likewise, the fermionic projection of the curvature is
\begin{align}\label{gfer_intermediate_nomc}
    &\left( d_{\mathfrak{L}} \mathfrak{L} \right) \big\vert_{\mathfrak{g}_F} \nonumber \\
    &\quad = \left( z + \frac{\ell}{1 - \frac{\kay}{\hay} } \right) \sqrt{ \frac{ \ell \left( 1 - \frac{\kay}{\hay} \right) }{z^2 - 1 } } \left( \partial_+ j_{-}^{(1)} - \partial_- j_+^{(1)} \right) \nonumber \\
    &\qquad + \left( z - \frac{\ell}{1 + \frac{\kay}{\hay} } \right) \sqrt{ \frac{ \ell \left( 1 + \frac{\kay}{\hay} \right) }{ z^2 - 1 } } \left( \partial_+ j_{-}^{(3)} - \partial_- j_+^{(3)} \right) \nonumber \\
    &\qquad + \left( z + \frac{\ell}{1 - \frac{\kay}{\hay} } \right) \sqrt{ \frac{ \ell \left( 1 - \frac{\kay}{\hay} \right) }{z^2 - 1 } } \left( [ j_+^{(0)} , j_-^{(1)} ] + [ j_+^{(1)} , j_-^{(0)} ] \right) \nonumber \\
    &\qquad + \left( z - \frac{\ell}{1 + \frac{\kay}{\hay} } \right) \sqrt{ \frac{ \ell \left( 1 + \frac{\kay}{\hay} \right) }{ z^2 - 1 } } \left( [ j_+^{(0)} , j_-^{(3)}] + [ j_+^{(3)} , j_-^{(0)} ] \right) \nonumber \\
    &\qquad + \ell \frac{z^2 + 1}{z^2 - 1} \left( z + \frac{\ell}{1 - \frac{\kay}{\hay} } \right) \sqrt{ \frac{ \ell \left( 1 - \frac{\kay}{\hay} \right) }{z^2 - 1 } }  \left( [ j_+^{(1)} , j_-^{(2)} ] + [ j_+^{(2)} , j_-^{(1)} ] \right) \nonumber \\
    &\qquad + \ell \frac{z^2 + 1}{z^2 - 1} \left( z - \frac{\ell}{1 + \frac{\kay}{\hay} } \right) \sqrt{ \frac{ \ell \left( 1 + \frac{\kay}{\hay} \right) }{ z^2 - 1 } }  \left( [ j_+^{(2)} , j_-^{(3)} ] + [ j_+^{(3)} , j_-^{(2)} ]  \right) \nonumber \\
    &\qquad + \left( \frac{\kay}{\hay} - \frac{2 \ell z}{z^2 - 1 } \right) \left( z + \frac{\ell}{1 - \frac{\kay}{\hay} } \right) \sqrt{ \frac{ \ell \left( 1 - \frac{\kay}{\hay} \right) }{z^2 - 1 } }  \left( [ \mathfrak{J}_+^{(2)} , j_-^{(1)} ] - [ j_+^{(1)} , \mathfrak{J}_-^{(2)} ] \right) \nonumber \\
    &\qquad + \left( \frac{\kay}{\hay} - \frac{2 \ell z}{z^2 - 1 } \right) \left( z - \frac{\ell}{1 + \frac{\kay}{\hay} } \right) \sqrt{ \frac{ \ell \left( 1 + \frac{\kay}{\hay} \right) }{ z^2 - 1 } } \left( [ \mathfrak{J}_+^{(2)} , j_-^{(3)} ] - [ j_+^{(3)} , \mathfrak{J}_-^{(2)} ] \right) \, .
\end{align}
Our task is to simplify the two projections (\ref{gbos_intermediate_no_mc}) and (\ref{gfer_intermediate_nomc}) as much as possible using the Maurer-Cartan identities (\ref{sSSSM_four_maurer_cartan_nicer}), the $v_{\pm}$ equation of motion, and the assumption
\begin{align}\label{ellsq_relation}
    \ell^2 = 1 - \frac{\kay^2}{\hay^2} \, ,
\end{align}
which is required for classical integrability even in the undeformed sSSSM-WZ.

Let us begin by simplifying the bosonic part (\ref{gbos_intermediate_no_mc}). The terms involving partial derivatives $\partial_{\pm} j_{\mp}^{(2)}$ and $\partial_{\pm} \mathfrak{J}_{\mp}^{(2)}$ in the first line of the right side precisely combine with the commutator terms in the fourth line of the right side to form covariant derivatives $D_{\pm} j_{\mp}^{(2)}$ and $D_{\pm} \mathfrak{J}_{\mp}^{(2)}$. Then replacing the combinations $F_{+-}^{(0)}$ and $D_+ j_-^{(2)} - D_- j_+^{(2)}$ using the $\mathfrak{g}_0$ and $\mathfrak{g}_2$ projections of the Maurer-Cartan identity given in equation (\ref{sSSSM_four_maurer_cartan_nicer}), respectively, this is
\begin{align}\label{gbos_intermediate_no_mc_again}
    &\left( d_{\mathfrak{L}} \mathfrak{L} \right) \big\vert_{\mathfrak{g}_B} \nonumber \\
    &\quad \deq - [ j_+^{(1)} , j_-^{(3)} ] \!-\! [ j_+^{(2)} , j_-^{(2)} ] \!-\! [ j_+^{(3)} , j_-^{(1)} ] \!-\! \left( \frac{\kay}{\hay} \!-\! \frac{2 \ell z}{z^2 \!-\! 1 } \right) \left( D_- \mathfrak{J}_+^{(2)} \!+\! D_+ \mathfrak{J}_-^{(2)} \right) \nonumber \\
    &\qquad + \left( \ell^2  \left( \frac{z^2 \!+\! 1}{z^2 \!-\! 1} \right)^2 \!-\! \left( \frac{\kay}{\hay} \!-\! \frac{2 \ell z}{z^2 \!-\! 1 } \right)^2 \right) [ j_+^{(2)} , j_-^{(2)} ] \nonumber \\
    &\qquad + \frac{\ell}{z^2 \!-\! 1} \left( 1 \!-\! \frac{\ell^2}{1 \!-\! \frac{\kay}{\hay}} \!-\! 2 \ell z \!+\! \frac{\kay}{\hay} z^2 \right) [ j_+^{(1)} , j_-^{(1)} ] \nonumber \\
    &\qquad + \frac{\ell}{z^2 \!-\! 1} \left( \!-\! 1 \!+\! \frac{\ell^2}{1 \!+\! \frac{\kay}{\hay}} \!-\! 2 \ell z \!+\! \frac{\kay}{\hay} z^2 \right) [ j_+^{(3)} , j_-^{(3)} ]  \nonumber \\
    &\qquad + \left( z \!+\! \frac{\ell}{1 \!-\! \frac{\kay}{\hay} } \right) \sqrt{ \frac{ \ell \left( 1 \!-\! \frac{\kay}{\hay} \right) }{z^2 \!-\! 1 } } \left( z \!-\! \frac{\ell}{1 \!+\! \frac{\kay}{\hay} } \right) \sqrt{ \frac{ \ell \left( 1 \!+\! \frac{\kay}{\hay} \right) }{ z^2 \!-\! 1 } } \left( [ j_+^{(1)} , j_-^{(3)} ] \!+\! [ j_+^{(3)} , j_-^{(1)} ]  \right) \, .
\end{align}
We now apply the relation (\ref{ellsq_relation}) between the parameters, which simplifies many of the combinations appearing in (\ref{gbos_intermediate_no_mc_again}). When this constraint is obeyed, one finds
\begin{align}\label{gbos_intermediate_no_mc_three}
    \left( d_{\mathfrak{L}} \mathfrak{L} \right) \big\vert_{\mathfrak{g}_B} &\deq - \left( \frac{\kay}{\hay} - \frac{2 \ell z}{z^2 - 1 } \right) \left( D_- \mathfrak{J}_+^{(2)} + D_+ \mathfrak{J}_-^{(2)} + 2 \frac{\kay}{\hay} [ j_+^{(2)} , j_-^{(2)} ] \right) \nonumber \\
    &\qquad + \ell \left( \frac{\kay}{\hay} - \frac{2 \ell z}{z^2 - 1} \right) \left( [ j_+^{(3)} , j_-^{(3)} ] - [ j_+^{(1)} , j_-^{(1)} ] \right) \nonumber \\
    &\qquad - \frac{\kay}{\hay} \left( \frac{\kay}{\hay} -  \frac{2 \ell z}{z^2 - 1} \right) \left( [ j_+^{(1)} , j_-^{(3)} ] + [ j_+^{(3)} , j_-^{(1)} ] \right) \, .
\end{align}
Therefore, when the Maurer-Cartan identity (\ref{sSSSM_four_maurer_cartan_nicer}), the $v_{\pm}$ equation of motion, and the relation (\ref{ellsq_relation}) are satisfied, we find that the bosonic projection of the curvature $d_{\mathfrak{L}} \mathfrak{L}$ is
\begin{align}\label{bosonic_sSSSM_lax_final}
    \left( d_{\mathfrak{L}} \mathfrak{L} \right) \big\vert_{\mathfrak{g}_B} \deq - \left( \frac{\kay}{\hay} - \frac{2 \ell z}{z^2 - 1} \right) &\Bigg( D_- \mathfrak{J}_+^{(2)} + D_+ \mathfrak{J}_-^{(2)} + \frac{\kay}{\hay} \left( 2 [ j_+^{(2)} , j_-^{(2)} ] +  [ j_+^{(1)} , j_-^{(3)} ] + [ j_+^{(3)} , j_-^{(1)} ] \right) \nonumber \\
    &\qquad \qquad - \ell \left( [ j_+^{(3)} , j_-^{(3)} ] - [ j_+^{(1)} , j_-^{(1)} ] \right) \Bigg) \, .
\end{align}
Next we consider the fermionic part. Beginning from (\ref{gfer_intermediate_nomc}), we again combine terms involving partial derivatives with commutators to form covariant derivatives, which gives
\begin{align}\label{gfer_intermediate_nomc_two}
    &\left( d_{\mathfrak{L}} \mathfrak{L} \right) \big\vert_{\mathfrak{g}_F} \nonumber \\
    &\quad \deq \left( z + \frac{\ell}{1 - \frac{\kay}{\hay} } \right) \sqrt{ \frac{ \ell \left( 1 - \frac{\kay}{\hay} \right) }{z^2 - 1 } } \left( D_+ j_{-}^{(1)} - D_- j_+^{(1)} \right) \nonumber \\
    &\qquad + \left( z - \frac{\ell}{1 + \frac{\kay}{\hay} } \right) \sqrt{ \frac{ \ell \left( 1 + \frac{\kay}{\hay} \right) }{ z^2 - 1 } } \left( D_+ j_{-}^{(3)} - D_- j_+^{(3)} \right) \nonumber \\
    &\qquad + \ell \frac{z^2 + 1}{z^2 - 1} \left( z + \frac{\ell}{1 - \frac{\kay}{\hay} } \right) \sqrt{ \frac{ \ell \left( 1 - \frac{\kay}{\hay} \right) }{z^2 - 1 } }  \left( [ j_+^{(1)} , j_-^{(2)} ] + [ j_+^{(2)} , j_-^{(1)} ] \right) \nonumber \\
    &\qquad + \ell \frac{z^2 + 1}{z^2 - 1} \left( z - \frac{\ell}{1 + \frac{\kay}{\hay} } \right) \sqrt{ \frac{ \ell \left( 1 + \frac{\kay}{\hay} \right) }{ z^2 - 1 } }  \left( [ j_+^{(2)} , j_-^{(3)} ] + [ j_+^{(3)} , j_-^{(2)} ]  \right) \nonumber \\
    &\qquad + \left( \frac{\kay}{\hay} - \frac{2 \ell z}{z^2 - 1 } \right) \left( z + \frac{\ell}{1 - \frac{\kay}{\hay} } \right) \sqrt{ \frac{ \ell \left( 1 - \frac{\kay}{\hay} \right) }{z^2 - 1 } }  \left( [ \mathfrak{J}_+^{(2)} , j_-^{(1)} ] - [ j_+^{(1)} , \mathfrak{J}_-^{(2)} ] \right) \nonumber \\
    &\qquad + \left( \frac{\kay}{\hay} - \frac{2 \ell z}{z^2 - 1 } \right) \left( z - \frac{\ell}{1 + \frac{\kay}{\hay} } \right) \sqrt{ \frac{ \ell \left( 1 + \frac{\kay}{\hay} \right) }{ z^2 - 1 } } \left( [ \mathfrak{J}_+^{(2)} , j_-^{(3)} ] - [ j_+^{(3)} , \mathfrak{J}_-^{(2)} ] \right) \, .
\end{align}
We again apply the $\mathfrak{g}_1$ and $\mathfrak{g}_3$ projections of the Maurer-Cartan identity (\ref{sSSSM_four_maurer_cartan_nicer}) to eliminate these terms involving covariant derivatives. Finally, we use assumption (\ref{ellsq_relation}), which allows us to simplify some of the coefficients. After doing this and simplifying, one obtains
\begin{align}\label{fermion_final}
    \left( d_\mathfrak{L} \mathfrak{L} \right) \big\vert_{\mathfrak{g}_F} &= - \left( z + \frac{\ell}{1 - \frac{\kay}{\hay} } \right) \sqrt{ \frac{ \ell \left( 1 - \frac{\kay}{\hay} \right) }{z^2 - 1 } }  \left( [ j_+^{(2)} , j_-^{(3)} ] + [ j_+^{(3)} , j_-^{(2)} ] \right) \nonumber \\
    &\quad - \left( z - \frac{\ell}{1 + \frac{\kay}{\hay} } \right) \sqrt{ \frac{ \ell \left( 1 + \frac{\kay}{\hay} \right) }{ z^2 - 1 } } \left( [ j_+^{(1)} , j_-^{(2)} ] + [ j_+^{(2)} , j_-^{(1)} ] \right) \nonumber \\
    &\quad + \ell \frac{z^2 + 1}{z^2 - 1} \left( z + \frac{\ell}{1 - \frac{\kay}{\hay} } \right) \sqrt{ \frac{ \ell \left( 1 - \frac{\kay}{\hay} \right) }{z^2 - 1 } } \left( [ j_+^{(2)} , j_-^{(1)} ] + [ j_+^{(1)} , j_-^{(2)} \right) \nonumber \\
    &\quad + \ell \frac{z^2 + 1}{z^2 - 1} \left( z - \frac{\ell}{1 + \frac{\kay}{\hay} } \right) \sqrt{ \frac{ \ell \left( 1 + \frac{\kay}{\hay} \right) }{ z^2 - 1 } } \left( [ j_+^{(2)} , j_-^{(3)} ] + [ j_+^{(3)} , j_-^{(2)} ] \right) \nonumber \\
    &\quad + \left( \frac{\kay}{\hay} - \frac{2 \ell z}{z^2 - 1 } \right) \left( z + \frac{\ell}{1 - \frac{\kay}{\hay} } \right) \sqrt{ \frac{ \ell \left( 1 - \frac{\kay}{\hay} \right) }{z^2 - 1 } } \left( [ \mathfrak{J}_+^{(2)} , j_-^{(1)} ] + [ \mathfrak{J}_-^{(2)} , j_+^{(1)} ] \right) \nonumber \\
    &\quad  + \left( \frac{\kay}{\hay} - \frac{2 \ell z}{z^2 - 1 } \right) \left( z - \frac{\ell}{1 + \frac{\kay}{\hay} } \right) \sqrt{ \frac{ \ell \left( 1 + \frac{\kay}{\hay} \right) }{ z^2 - 1 } } \left( [ \mathfrak{J}_+^{(2)} , j_-^{(3)} ] + [ \mathfrak{J}_-^{(2)} , j_+^{(3)} \right) \, .
\end{align}

\allowdisplaybreaks

\bibliographystyle{utphys}
\bibliography{master}

\end{document}